\documentclass[12pt, draftclsnofoot, onecolumn]{IEEEtran}
\IEEEoverridecommandlockouts
\usepackage{cite}
\usepackage{amsmath,amssymb,amsfonts}
\usepackage{caption}
\usepackage{graphicx}
\usepackage{booktabs}
\usepackage{epstopdf}
\usepackage[tight,footnotesize]{subfigure}
\usepackage{color} 
\ifCLASSINFOpdf
\else
\fi
\allowdisplaybreaks[4]
\begin{document}
	\title{Hybrid Spherical- and Planar-Wave Channel Modeling and DCNN-powered Estimation for Terahertz Ultra-massive MIMO Systems}
	\author{Yuhang~Chen,
		Longfei~Yan,~and~Chong~Han,~\IEEEmembership{Member,~IEEE}
		\thanks{This paper was presented in part at the IEEE GLOBECOM, 2020~\cite{DCNN_GC}.}
		\thanks{Y. Chen, L. Yan and C. Han are with the Terahertz Wireless Communications (TWC) Laboratory, Shanghai Jiao Tong University, Shanghai 200240, China (e-mail: \{yuhang.chen, longfei.yan, chong.han\}@sjtu.edu.cn).
		}
	}
	\maketitle
	\thispagestyle{empty}
	\begin{abstract}
		
		The Terahertz band is envisioned to meet the demanding 100 Gbps data rates for 6G wireless communications. 
		Aiming at combating the distance limitation problem with low hardware-cost, ultra-massive MIMO with hybrid beamforming is promising. 
		However, relationships among wavelength, array size and antenna spacing give rise to the inaccuracy of planar-wave channel model~(PWM), while an enlarged channel matrix dimension leads to excessive parameters of applying spherical-wave channel model~(SWM).
		Moreover, due to the adoption of hybrid beamforming, channel estimation~(CE) needs to recover high-dimensional channels from severely compressed channel observation. 
		In this paper, a hybrid spherical- and planar-wave channel model~(HSPM) is investigated and proved to be accurate and efficient by adopting PWM within subarray and SWM among subarrays.
		Furthermore, a two-phase HSPM CE mechanism is developed. A deep convolutional-neural-network~(DCNN) is designed in the first phase for parameter estimation of reference subarrays, while geometric relationships of the remaining channel parameters between reference subarrays are leveraged to complete CE in the second phase. 
		Extensive numerical results demonstrate the HSPM is accurate at various communication distances, array sizes and carrier frequencies.
		The DCNN converges fast and achieves high accuracy with 5.2~dB improved normalized-mean-square-error compared to literature methods, and owns substantially low complexity. 
	\end{abstract}
	
	\columnsep 0.2in
	\section{Introduction}
	
	Owning ultra-broad multi-GHz bandwidth, Terahertz (THz) wireless communications are expected to meet the future demand of 100+ Gbps wireless data rates, which are therefore envisioned as an enabling technology for 6G communications~\cite{THz_demand}. 
	The enticingly high data rates of THz communications come at a price of restricted wireless communication distance, due to the high propagation losses induced by the severe spreading loss and atmospheric attenuation in the THz band.
	By generating beams with high beamforming gain to compensate for the path loss, ultra-massive MIMO (UM-MIMO) is promising and widely adopted in THz wireless systems~\cite{THz_demand,ref_combat_dist,ref_hybrid_magazine} to address the distance limitation problem.

	On one hand, as the fundamental basis of analyzing and designing THz UM-MIMO systems to achieve the promised capability, accurate channel modeling is critical. However, although the traditional planar-wave channel model (PWM) in the micro- and millimeter-wave frequency bands requires only a small number of channel parameters proportional to the number of multi-path to determine the channel, it becomes inaccurate to THz UM-MIMO systems, especially when the carrier frequency and array size increase, and the spherical wavefront is demanded to accurately analyze the propagation of THz waves~\cite{ref_hybrid_magazine,ref_spherical_fronts,ref_2level}.
	Nevertheless, an exponentially increasing number of parameters proportional to the massive number of antennas in THz UM-MIMO systems is foreseen by directly applying the spherical-wave channel model (SWM).
	
	On the other hand, retrieval of accurate channel-state-information (CSI) by channel estimation~(CE) is critical in establishing communication links. However, distinctive challenges are encountered for CE in THz UM-MIMO systems.
	First, since CE is closely related to the channel model, an efficient CE scheme requires precisely characterizing the spherical-wave propagation demand.
	Second, being a cost-effective alternative of conventional fully digital MIMO structure, the hybrid UM-MIMO structure that employs a much smaller number of RF chains to control large size of antenna arrays has drawn enormous interest to THz communications~\cite{ref_hybrid_magazine,DAoSA}. The received signal is compressed to the dimension of the number of RF-chains, leading that CE requires to recover the high dimensional channel with severely compressed channel observations.
	
	To this end, modeling the THz UM-MIMO channel with high accuracy and a small number of parameters, and effective CE are stringently needed in THz UM-MIMO systems.
	
	\subsection{Related Work} 
	\subsubsection{Channel Modeling}
	There are mainly two selections of MIMO channel models in the literature, namely, the SWM and the PWM~\cite{ref_SW_PW_Modeling,ref_spherical_fronts}. 
	The SWM takes the spherical radio-wave front into consideration, which is the most accurate model of characterizing the radio-wave propagation~\cite{ref_SW_PW_Modeling}. 
	To determine the SWM, $2N_p(N_t\times N_r)$ parameters are used, where $N_t$, $N_r$ and $N_p$ describe the number of antennas at the transmitter (Tx) and receiver (Rx), and the number of multi-path, whose typical values are $N_t=N_r=1024$ and $N_p\leq 10$ in the THz UM-MIMO systems~\cite{DAoSA}.
	By contrast, the PWM approximates the radio-wave front as a plane, which remains accurate when the array size is relatively small with a large communication distance~\cite{ref_SW_PW_Modeling}. 
	$6N_p$ channel parameters are explored to determine the PWM, which include the azimuth and elevation angles of departure as well as arrival, the amplitude of the path gain and communication distance.

	However, SWM processes high complexity, while PWM suffers from low accuracy in modeling the THz UM-MIMO channels, respectively.
	Particularly, since the SWM requires a large number of channel parameters to determine the channel matrix, most of the research efforts focus on the PWM~\cite{ref_3D_channel,ref_Channel_smart_radio}.
	Nevertheless, as carrier frequency and array size increase, the approximation error caused by the PWM becomes non-negligible to THz UM-MIMO systems~\cite{ref_hybrid_magazine}. 
	Motivated by this, channel models by combining the PWM and SWM are studied in~\cite{ref_2level,ref_WSMS}. Particularly, by taking the subarray and a unit, and considering that common reflectors are shared among subarrays, \cite{ref_2level} and~\cite{ref_WSMS} use the same channel parameters including the direction-of-departure (DoD), direction-of-arrival (DoA), and the amplitude of channel gain to describe the channels.
	A phase-shift coefficient is introduced to address the effect of spherical-wave propagation among subarrays.
	However, practical spherical-wave propagation results in \textit{parameter-shift}, where the channel parameters vary over different antennas in the large-size array~\cite{ref_CM_NSWM,ref_CM_mmW}, the channel models in~\cite{ref_2level,ref_WSMS} are still inaccurate.
	Although the idea of taking the subarray as a unit to model the channel has been studied in the distributed MIMO systems~\cite{ref_distrubited_MIMO1, ref_distrubited_MIMO2}, these systems consider very long geometric distances up to tens of meters between subarrays. Different subarrays do not share common reflectors, which is different from the systems in~\cite{ref_2level} and~\cite{ref_WSMS}.

	\subsubsection{Channel Estimation}
	In the literature, various solutions including the conventional \textit{on-grid}~\cite{OMP,AMP,ref_yuan_CE} and \textit{off-grid}~\cite{ref_ESPRIT_OFDM,AoSA_MUSIC,ref_sp_ce_hu,ref_EM_Bayesian},
	and the recently proposed deep-learning~(DL)-based solutions~\cite{LDAMP,ref_GM_LAMP,ref_sparse_CE_DL,ref_dl_CE, mmWave_CE_CNN,ref_CE_GAN} have been studied for UM-MIMO CE.
	The on-grid solutions consider the directions of the propagation paths are taken from fixed spatial grids, which include the compressive-sensing (CS) solutions such as the orthogonal matching pursuit (OMP)~\cite{OMP}, the approximate matching pursuit (AMP)~\cite{AMP} and others~\cite{ref_yuan_CE}.
	Since the angles of propagation path are continuous-valued in practice, these solutions suffer from limited accuracy due to grid-mismatch, namely, the power leakage effect~\cite{Super_ChenHu}.
	By contrast, the off-grid solutions eliminate the on-grid assumption for improved accuracy. Particularly, the subspace-based estimation with rotational invariance technique~\cite{ref_ESPRIT_OFDM} and AoSA-MUSIC~\cite{AoSA_MUSIC} perform CE by eigenvalue-decomposition. The grid-refinement solution~\cite{ref_sp_ce_hu} increases the grid-resolution of the CS solutions. 
	The expectation-maximization~\cite{ref_EM_Bayesian} exploits the relation between the channel parameters and the received signal for parameter estimation. 
	However, the high accuracy of the off-grid solutions comes at a cost of higher complexity. 
	Moreover, traditional on-grid and off-grid solutions cannot achieve satisfactory estimation accuracy when the channel possesses the spherical-wave propagation attribute~\cite{DCNN_GC}.

	Recently, with the rapid advancement of DL techniques for wireless communications, DL-based
	CE is explored to excavate the inherent characteristic of the channel, which achieves improved CE performance and is a good candidate to solve the UM-MIMO CE problem~\cite{ LDAMP,ref_GM_LAMP,ref_sparse_CE_DL,ref_dl_CE, mmWave_CE_CNN,ref_CE_GAN}.
	The DL-based solutions can be divided into model-driven~\cite{LDAMP,ref_GM_LAMP}, and data-driven methods~\cite{ref_sparse_CE_DL,ref_dl_CE,mmWave_CE_CNN,ref_CE_GAN}. 
	The model-driven method is designed based on traditional iterative algorithms, in which each layer of the network represents an iteration. 
	Despite a faster convergence rate compared to the original iterative algorithm, the performance of the model-driven method is highly dependent on the original iterative algorithm~\cite{LDAMP,ref_GM_LAMP}.  
	By contrast, the data-driven method is model-independent, which is applicable in various scenarios~\cite{ref_sparse_CE_DL,ref_dl_CE,mmWave_CE_CNN,ref_CE_GAN}. 
	Moreover, the data-driven method obtains high precision with proper network training.
	Nevertheless, owing to the extremely large channel dimension of THz UM-MIMO systems, these solutions suffer from high complexity.
	In addition, all the existing DL-based solutions do not address the spherical-wave propagation property in the THz UM-MIMO systems.
	Therefore, more effective DL-based CE solutions with low complexity are needed for THz UM-MIMO systems.

	\subsection{Contributions}
	In this paper, by considering a generalized THz UM-MIMO structure, we address the challenges in channel modeling and CE as mentioned above.
    First, we analytically derive the closed-form results of the approximation error between the PWM and SWM for two-dimensional~(2D) planar arrays. 
	Then, we investigate an analytical hybrid spherical- and planar-wave channel model (HSPM) accounting for the spherical-wave propagation and parameter-shift effect among subarrays.
	Based on that, comparisons among three different channel models are further conducted, which prove the enhancement on modeling accuracy and reduction on the number of channel parameters for the HSPM, e.g., from $2N_p(N_t\times N_r)$ to $6N_p$.
	A two-phase CE mechanism is further proposed, which first directly estimates the $6N_p$ channel parameters between reference subarrays based on a designed deep convolutional-neural-network~(DCNN) network. 
	Then, the geometric relationships between channel parameters of the remaining subarrays and the reference subarray are derived to construct the full channel matrix. 
	
	In our prior and shorter version of this work~\cite{DCNN_GC}, we adopt the channel model in~\cite{ref_2level} and develop a DCNN CE solution. In this work, we derive the closed-form approximation error for the 2D planar array between the PWM and SWM, and adopt a new HSPM channel model. We also develop a two-phase DCNN CE mechanism for HSPM, with substantially more performance evaluation and analysis. 
	The main contributions of this work are summarized as follows.

	\begin{itemize}
		\item \textbf{We investigate the HSPM and analytically derive the approximation errors in closed-form between different channel models.}
			We consider the 2D planar array and derive the approximation error between the PWM and SWM. 
			By accounting for the spherical-wave propagation among subarrays and the parameter-shift effect, we investigate the HSPM for THz UM-MIMO systems. 
			Comparisons among different channel models confirm that compared to PWM, HSPM uses slightly larger $N_p(1+5K_tK_r)$ parameters to determine the THz channel with greatly improved modeling accuracy, where $K_t$ and $K_r$ denote the number of subarrays at Tx and Rx, respectively.

		\item \textbf{We develop the two-phase CE mechanism by accounting for the features of the HSPM.}
		In the first phase, we develop a DCNN network to directly estimate $6N_p$ channel parameters, including the azimuth and elevation angles of departure and arrival, the amplitude of path gain, and communication distances between the reference subarrays, to achieve reduced complexity and high parameter estimation accuracy.
		In the second phase, relations of parameters between the reference subarray and remaining subarrays are explored by geometric relationships. Finally, the channel matrix is reconstructed to complete the CE process. 
		
		\item \textbf{We use a ray-tracing tool to generate the simulated channel data, which are used to provide extensive numerical results and evaluate the performance of the HSPM and two-phase CE mechanism, respectively.}
		Results demonstrate that the HSPM remains accurate with different communication distances, array sizes and carrier frequencies, 
		The designed DCNN converges fast and achieves high-resolution parameter estimation and CE with substantially reduced complexity. 
	\end{itemize}
	
	The remainder of the paper is organized as follows. In Sec.~\ref{sec_UM-MIMO System Model}, the system and signal models of THz UM-MIMO systems are investigated. In Sec.~\ref{sec_Hybrid Spherical- and Planar-wave Channel Model}, we introduce the SWM and PWM. Based on the derived approximation error of the PWM, the HSPM is investigated. 
	The two-phase CE mechanism is develpoed in Sec.~\ref{sec_Two-phase Channel Estimation}. 
	After an in-depth analysis and numerical evaluation of the proposed HSPM and CE mechanism in Sec.~\ref{sec_Performance_Evaluation}, the paper is summarized in Sec.~\ref{sec_Conclusion}. 
	
	\textbf{Notation}: 
	$a$ is a scalar.
	$\mathbf{a}$ denotes a vector. 
	$\mathbf{A}$ represents a matrix. 
	$\mathbf{A}[m,n]$ stands for the element at the $m^{\rm th}$ row and $n^{\rm th }$ column in $\mathbf{A}$.
	$\mathbb{C}^{M\times N}$ depicts the set of $M\times N$-dimensional complex-valued matrices. 
	$(\cdot)^{\mathrm{T}}$ defines transpose.
	$(\cdot)^{\mathrm{H}}$ refers to conjugate transpose.
	$\mathrm{exp}{\{\cdot \}}$ defines the exponential function of $\mathrm{e}$.
	$\mathbb{E}\{\cdot\}$ describes the expectation. 
	$\mathbf{I}_{N}$ defines an $N$ dimensional identity matrix.
	$\textrm{Re}[\cdot]$ is to take the real number. 
	$\textrm{Im}[\cdot]$ refers to take the imaginary number.
	$*$ represents convolution operation. 
	$|\cdot|$ denotes the absolute value.
	$\Vert\cdot\Vert$ stands for the $2$-norm.
	$\Vert\cdot\Vert_{\rm{F}}$ defines the Frobenius norm.

	\section{UM-MIMO System Model} \label{sec_UM-MIMO System Model}
	In this section, we introduce the system model and the received signal model for THz UM-MIMO systems. 
	
	\subsection{System Model}
	
	\begin{figure}[t]
		\centering
		{\includegraphics[width= 0.95\textwidth]{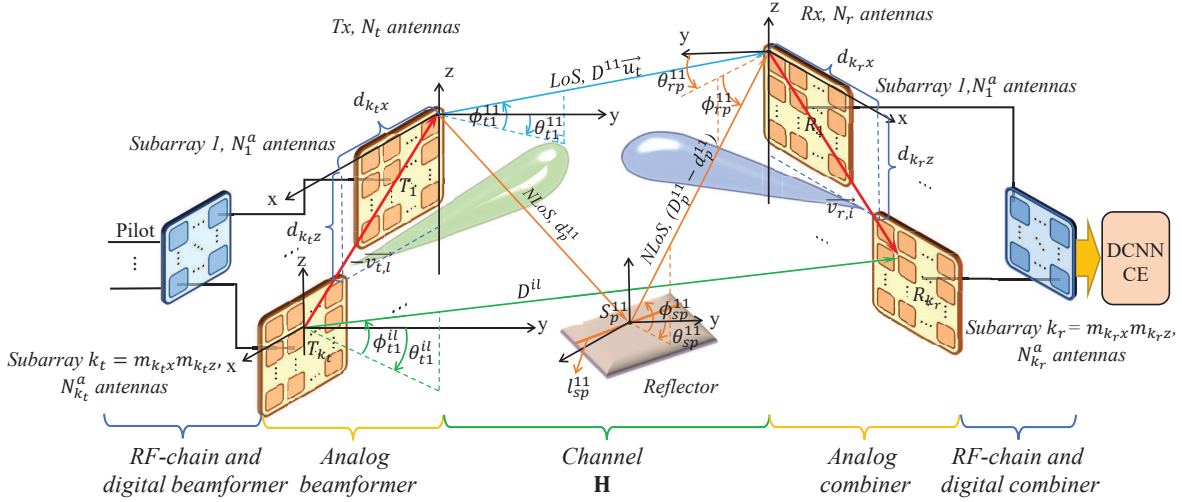}}
		\caption{THz UM-MIMO system model.} 
		\label{fig_system_model}
		\vspace{-6mm}
	\end{figure}

	As illustrated in Fig.~\ref{fig_system_model}, we consider the generalized THz UM-MIMO systems are equipped at Tx and Rx. There are $N_t$ transmitted antennas and $N_r$ received antennas, which are divided into $K_t$ and $K_r$ subarrays at Tx and Rx, respectively. 
	We select the first subarray at Tx and Rx, denoted by $T_1$ and $R_1$, as the reference subarrays, respectively. For each subarray, the first antenna is chosen as the reference antenna.
	The antenna spacing $d = \lambda/2$ within a subarray, where $\lambda$ denotes the carrier wavelength. The subarray spacing is arbitrarily multiple of half-wavelength, which stands for the general setting of the THz UM-MIMO~\cite{ref_2level,ref_WSMS}. 
	Particularly, the distances between $T_1$ and the $k_t^{\rm th}$ subarray $T_{k_t}$ at Tx alone x-axis and z-axis are denoted as $d_{k_tx} = m_{k_tx}d$ and $d_{k_tz}=m_{k_tz}d$, respectively, where $k_t = 1, ..., K_t$, the integers $m_{k_tx},m_{k_tz}\geq 1$.
	Similarly, $d_{k_rx}=m_{k_rx}d$ and $d_{k_rz}=m_{k_rz}d$ stand for the distances between $R_1$ and the $k_r^{\rm th}$ subarray $R_{k_r}$ at Rx on x-axis and z-axis, respectively, with $k_r = 1,...,K_r$ and $m_{rx},m_{rz}\geq 1$.
	In addition,
	$m_{tx} = 0,1,...,M_{K_tx}-1$,
	$m_{tz} = 0,1,...,M_{K_tz}-1$,
	$m_{rx} = 0,1,...,M_{K_rx}-1$,
	$m_{rz} = 0,1,...,M_{K_rz}-1$, 
	where $M_{K_tx}$ and $M_{K_tz}$ represent the number of subarrays along x-axis and z-axis at Tx, with $K_t=M_{K_tx}M_{K_tz}$, while
	$M_{K_rx}$ and $M_{K_rz}$ stand for the number of subarrays along x-axis and z-axis at Rx, with $K_r=M_{K_rx}M_{K_rz}$, respectively. 

	Each subarray connects to one RF-chain, to perform transmitted analog beamforming and received analog combining. 
	The analog beamforming matrix $\mathbf{F}_{\rm RF}\in\mathbb{C}^{N_t\times K_t}$ and combining matrix $\mathbf{W}_{\rm RF}\in\mathbb{C}^{N_r\times K_r}$ hold the block diagonal structure, in which $\mathbf{W}_{\rm RF}$ is expressed as
	\begin{equation}\label{equ_block_diagonal}
	\mathbf{W}_{\rm RF}=\left[\begin{matrix}
	\mathbf{w}_{1}&0&\ldots&0\\
	0&\mathbf{w}_2&\ldots&0\\
	\vdots&\vdots&\ddots&\vdots\\
	0&0&\ldots&\mathbf{w}_{K_r}
	\end{matrix}\right],
	\end{equation}
	where $\mathbf{w}_{k_r}=[\omega_{1,k_r},\ldots,\omega_{N_{k_r}^a,k_r}]^{\mathrm{T}}\in\mathbb{C}^{N_{k_r}^a\times 1} $ represents the analog combining vector of the ${k_r}^{\mathrm{th}}$ RF-chain at Rx, with $k_r=\text{1},\ldots,K_r$, and ${N_{k_r}^a}$ denotes the number of antennas on the ${k_r}^{\mathrm{{th}}}$ subarray, with $N_r = \sum_{k_r=1}^{K_r}N^a_{k_r}$.
	Since analog beamforming and combining are implemented by phase shifters, each element in $\mathbf{w}_{k_r}$ satisfies the constant module constraint as
	\begin{equation}\label{equ_phase_shift_coefficient}
	\omega_{n_{k_r}^a,k}=
	\left(1/\sqrt{N_r}\right)\mathrm{e}^{j2\pi\tilde{\omega}_{n_{k_r}^a,k_r}},
	\end{equation}
	where $n_{k_r}^a=1,\ldots,N_{k_r}^a$, the phase shift coefficient satisfies $0\leq\tilde{\omega}_{n_{k_r}^a,k_r}\leq 1$. The transmitted beamforming matrix $\mathbf{F}_{\rm RF}$ has a similar block diagonal form as~\eqref{equ_block_diagonal}. The parameters of the THz UM-MIMO communication systems in Fig.~\ref{fig_system_model} are summarized in TABLE~\ref{Table_parameter}.
	
	\begin{table}[t]
		\centering
		\caption{Parameters of the THz UM-MIMO communication system.}
		\begin{tabular}{ccc}
			\toprule
			\textbf{Parameter}&\textbf{Symbol} & \textbf{Unit}\\
			\midrule
			Number of antennas at Tx &$N_t$ & 1\\
			Number of antennas at Rx &$N_r$&1 \\
			Number of RF-chains at Tx&$K_t$&1 \\
			Number of RF-chains at Rx&$K_r$&1 \\
			Distance between $T_1$ and $T_{k_t}$ alone x-axis& $d_{k_tx}$&m\\
			Distance between $T_1$ and $T_{k_t}$ alone z-axis& $d_{k_tz}$&m\\
			Distance between $R_1$ and $R_{k_r}$ alone x-axis& $d_{k_rx}$&m\\
			Distance between $R_1$ and $R_{k_r}$ alone z-axis& $d_{k_rz}$&m\\
			Distance between the reference antennas at Tx and Rx& $D^{11}$&m\\ 
			Distance between the reference antennas at Tx and the reflector of the $p^{\rm th}$ path & $d_p^{11}$&m\\
			Distance between the $l^{\rm th}$ Tx antenna to the $i^{\rm th}$ Rx antenna of the $p^{\rm th}$ path & $D^{il}_p$&m\\
			DoD pair of the LoS path between the reference antennas & $(\theta_{t1}^{11},\phi_{t1}^{11})$&rad\\
			DoA pair of the $p^{\rm th}$ path between the reference antennas & $(\theta_{rp}^{11},\phi_{rp}^{11})$&rad\\
			Angle pair of the $p^{\rm th}$ path between the reflector and the reference received antenna & $(\theta_{sp}^{11},\phi_{sp}^{11})$&rad\\
			DoD pair of the LoS path between the $l^{\rm th}$ Tx and $i^{\rm th}$ Rx antennas & $(\theta_{tp}^{il},\phi_{tp}^{il})$&rad\\
			Vector direct from the reference antenna to the $ l^{\rm th}$ antenna at Tx& $\overrightarrow{v_{t,l}} $&\\
			Vector direct from the reference antenna at Tx to the $ i^{\rm th}$ antenna at Rx& $\overrightarrow{u_{t}}$&\\
			Vector direct from the reference antenna to the $ i^{\rm th}$ antenna at Rx& $\overrightarrow{v_{r,i}}$ &\\
			Intersection line between the reflector and the plane of incoming and reflected rays &$l_{sp}^{11}$&\\
			\bottomrule
		\end{tabular}
		\vspace{-6mm}
		\label{Table_parameter}
	\end{table}
	
	\subsection{Received Signal}
	
	To overcome the beam misalignment problem in the THz band, a beam training procedure is required. Both Tx and Rx generate beams owning high beamforming gains to compensate the huge path loss and transmit pilot signal.
	Each beam is generated by a codeword stored in beam codebook~\cite{OMP}, which is constructed by changing the value of the phase shift coefficient in~\eqref{equ_phase_shift_coefficient}. 
	After Tx and Rx scan all the beam combinations, the received signal is constructed for CE. 
	At Tx, the pilot signal passes the digital and the analog beamformers to the channel. The received signal is combined by the analog and the digital combiners at Rx. 
	We denote the ${c}^{\rm{th}}$ codeword at Tx and Rx as $\mathbf{F}_{c} = \mathbf{F}_{{\rm{RF}}, c}\mathbf{F}_{{\rm {BB}}, c}\in\mathbb{C}^{N_t \times N_s }$ and $\mathbf{W}_{{c}}= \mathbf{W}_{{{\rm {RF}}}, c}\mathbf{W}_{{\rm {BB}}, c}\in\mathbb{C}^{N_r \times N_s}$, respectively, where $c = 1,...,C$ represents the codeword index, and $N_s\leq {\rm max}\{K_t,K_r\}$ describes the number of data streams. 
	Moreover, $\mathbf{F}_{{\rm {BB}}, c}\in\mathbb{C}^{K_t \times N_s}$ and $\mathbf{W}_{{\rm {BB}}, c}\in\mathbb{C}^{K_r\times N_s }$ denote the digital beamformer and combiner, respectively. 
	$\mathbf{F}_{{\rm {RF}}, c}\in\mathbb{C}^{N_t \times K_t}$ and $\mathbf{W}_{{\rm {RF}}, c}\in\mathbb{C}^{N_r \times K_r }$ represent the analog beamformer and combiner, respectively. 
	We use $\mathbf{S}\in\mathbb{C}^{N_s\times T}$ to represent the transmitted pilot signal, with $\mathbf{S}\mathbf{S}^{\mathrm{H}}=\mathbf{I}_{N_s}$, $T$ denotes the length of the pilot. 
	By denoting the THz UM-MIMO channel as $\mathbf{H} \in\mathbb{C}^{N_r\times N_t}$, the received signal $\mathbf{y}_{c,c}\in\mathbb{C}^{N_r\times T}$ is represented as
		\begin{equation}\label{Received_signal1}
		\begin{split}
		\overline{\mathbf{y}}_{c,c}&=\mathbf{W}_c^{\mathrm{H}}\mathbf{H}\mathbf{F}_c\mathbf{S}+\mathbf{W}_c^{\mathrm{H}}\overline{\mathbf{N}}_c,
		\end{split}
		\end{equation} 
		where $\overline{\mathbf{N}}_c\in\mathbb{C}^{N_r\times T}$ refers to the received complex additive white Gaussian noise (AWGN).
	After that, we multiply $\mathbf{S}^{\rm{H}}$ to the $\mathbf{y}_{c,c}$ to represent the matched filtering and obtain $\mathbf{y}_{c,c}\in\mathbb{C}^{N_s\times N_s}$
	\begin{equation}\label{Received_signal}
	\begin{split}
	\mathbf{y}_{c,c}&=\mathbf{W}_c^{\mathrm{H}}\mathbf{H}\mathbf{F}_c+\mathbf{W}_c^{\mathrm{H}}\mathbf{N}_c,
	\end{split}
	\end{equation} 
	where $\mathbf{N}_c=\overline{\mathbf{N}}_c\mathbf{S}^{\rm H}\in\mathbb{C}^{N_s\times N_s}$ stands for the modified noise. 
	
	By varying the beam codewords at both Tx and Rx to transverse all beam combinations, $C^2$ received signals as~\eqref{Received_signal} are acquired, which are contrasted together to obtain $\mathbf{Y}\in\mathbb{C}^{N_sC\times N_sC}$ as
	\begin{equation}\label{equ_Channel_observation}
	\begin{split}
	\mathbf{Y}&=\left[\begin{matrix}
	\mathbf{y}_{1,1}&\ldots&\mathbf{y}_{1,C}\\\vdots &\ddots&\vdots\\
	\mathbf{y}_{C,1}&\ldots&\mathbf{y}_{C,C}
	\end{matrix}\right]=\overline{\mathbf{W}}^{\mathrm{H}}\mathbf{H}\overline{\mathbf{F}}+\mathbf{N},
	\end{split}
	\end{equation}
	where $\overline{\mathbf{F}}=[\mathbf{F}_1,\ldots,\mathbf{F}_C]\in\mathbb{C}^{N_t\times N_sC}$ and $\overline{\mathbf{W}}=[\mathbf{W}_1,\ldots,\mathbf{W}_C]\in\mathbb{C}^{N_r\times N_sC}$, respectively, and $\mathbf{N}\in\mathbb{C}^{N_sC\times N_sC}$ denotes the stacked noise. 
	CE refers to estimating the channel matrix $\mathbf{H}$ based on the received signal $\mathbf{Y}$ in~\eqref{equ_Channel_observation}.
	However, due to the massive number of antennas in THz UM-MIMO systems, directly modeling and estimating the channel matrix suffers from high complexity. 
	
	\section{Hybrid Spherical- and Planar-wave Channel Model}\label{sec_Hybrid Spherical- and Planar-wave Channel Model}
	
	In this section, we first introduce the SWM and PWM. Then, we evaluate the approximation error of the PWM to analyze its applicability to THz UM-MIMO systems. Finally, the HSPM is considered, which achieves high accuracy and owns a reduced number of channel parameters. 
	
	\subsection{Spherical-wave Channel Model}\label{subsec_SWM}
	The	SWM is the most accurate model by individually calculating the channel responses of all antenna pairs between Tx and Rx. 
	As shown in Fig.~\ref{fig_system_model}, the 2D planar arrays are deployed at Tx and Rx, where $D^{il}$ denotes the communication distance from the $l^{\rm th}$ transmitted antenna to the $i^{\rm th}$ received antenna, with $i = 1,\ldots, N_t$ and $l = 1,\ldots, N_r$. 
	In the SWM, the complex path gain from the $l^{\rm th}$ transmitted antenna to the $i^{\rm th}$ received antenna, denoted by $\alpha^{il}$, is represented as
	\begin{equation}
	\label{equ_gain}
	\alpha^{il} = \lvert\alpha^{il}\rvert e^{-j\frac{2\pi}{\lambda}D^{il}}.
	\end{equation}
	
	The channel response between the $l^{\rm th}$ transmitted and $i^{\rm th}$ received antennas is expressed as
	\begin{equation}\label{equ_SW_channel}
	\mathbf{H}_{\rm S}[i,l]=\Sigma_{p=1}^{N_p}\lvert\alpha^{il}_p\rvert e^{-j\frac{2\pi}{\lambda}D^{il}_p},
	\end{equation} 
	where $\mathbf{H}_{\rm S}$ denotes the $N_r\times N_t$-dimensional spherical-wave channel matrix, 
	$p = 1,..., N_p$ indexes the propagation paths, 
	$p=1$ denotes the line-of sight (LoS) path, while $p>1$ represents non-line-of-sight (NLoS) paths.
	$\mathbf{H}_{\rm S}$ in~\eqref{equ_SW_channel} is dependent on the parameter set $\mathcal{P}_{S} = \{|\alpha^{il}_p|, D^{il}_p\}$ containing $ 2N_p(N_t\times N_r)$ elements, which increase exponentially with the number of antennas.

	\subsection{Planar-wave Channel Model}\label{subsec_PWM Method}
	The PWM is an approximation of the SWM when the array size is far less than the communication distance. 
	In particular, the signal transmission is approximated as parallel and the wavefront is analyzed as a plane. 
	As illustrated in Fig.~\ref{fig_system_model}, $(\theta_{tp}^{il}, \phi_{tp}^{il})$, $(\theta_{rp}^{il}, \phi_{rp}^{il})$ represent the DoD and DoA pairs for the $p^{\rm th}$ path between the $l^{\rm th}$ transmitted and the $i^{\rm th}$ received antennas, in which $\theta$ and $\phi$ denote the azimuth and elevation angles, respectively. 
	We consider the $l^{\rm th}$ transmitted antenna is located in the $k_t^{{\rm th}}$ subarray, and use the index pair $(n_{k_tx}, n_{k_tz})$ to denote the position of this antenna in the subarray, where $n_{k_tx} = 0,1,...,N_{k_tx}^{a}-1$, $n_{k_tz} = 0,1,...,N_{k_tz}^{a}-1$, and $N_{k_tx}^{a}$ and $N_{k_tz}^{a}$ represent the number of antennas of the $k_t^{{\rm th}}$ subarray along x-axis and z-axis, respectively. 
	Similarly, the index pair $(n_{k_rx}, n_{k_rz})$ determines the position of the $i^{\rm th}$ received antenna in the $k_r^{{\rm th}}$ subarray, where $n_{k_rx} = 0,1,...,N_{k_rx}^{a}-1$, $n_{k_rz} = 0,1,...,N_{k_rz}^{a}-1$, and $N_{k_rx}^{a}$ and $N_{k_rz}^{a}$ represent the number of antennas of the $k_r^{{\rm th}}$ subarray along x-axis and z-axis, respectively.
	
	By considering the plane-wave transmission, $D^{il}$ is approximated as~\cite{ref_2level}
	\begin{equation}
	D^{il}= D^{11}\!+\Delta D^{il}, 
	\end{equation}
	where $\Delta D^{il} =d( \psi_t-\psi_r)$, and 
	$\psi_t = (m_{tx}+n_{tx}){\rm sin}\theta_{tp}^{11}{\rm cos}\phi_{tp}^{11}-(m_{tz}+n_{tz}){\rm sin}\phi_{tp}^{11}$,
	$\psi_r = (m_{rx}+n_{rx}){\rm sin}\theta_{rp}^{11}{\rm cos}\phi_{rp}^{11}-(m_{rz}+n_{rz}){\rm sin}\phi_{rp}^{11}$ according to the geometric relationships in Fig.~\ref{fig_system_model}.
	Since $\Delta D^{il}$ is far less than $D^{11}$, $|\alpha^{il}|$ is considered approximately the same. The complex path gain between the $l^{\rm th}$ transmitted antenna and the $i^{\rm th}$ received antenna is approximated as~\cite{ref_SW_analysis}
	\begin{equation}\label{equ_approx_gain}
	\alpha^{il}\approx|\alpha^{11}|e^{-j\frac{2\pi}{\lambda} D^{il}} = |\alpha^{11}| e^{-j\frac{2\pi}{\lambda} D^{11}} e^{-j\frac{2\pi}{\lambda}\Delta D^{il}}.
	\end{equation}	
	Therefore, the planar-wave channel response between the $l^{\rm th}$ transmitted and $i^{\rm th}$ received antennas is represented as~\cite{ref_SW_analysis}
	\begin{equation}
	\begin{split}
	\mathbf{H}_{\rm P}[i,l]=&\Sigma_{p=1}^{N_p}\lvert\alpha^{11}_p\rvert e^{-j\frac{2\pi}{\lambda}D^{\!11}_p}e^{-j\frac{2\pi d}{\lambda}
		(\psi_t-\psi_r)}. 
	\end{split}
	\label{channel_planar}
	\end{equation}
	
	The planar-wave channel matrix in~\eqref{channel_planar} can be further arranged in a compact form as
	\begin{equation}\label{equ_P_model}
	\mathbf{H}_{\rm P}=\Sigma_{p=1}^{N_p}\lvert\alpha^{11}_p\rvert e^{-j\frac{2\pi}{\lambda}D^{11}_p}
	\mathbf{a}_{rp}(\theta_{rp}^{11},\phi_{rp}^{11})\mathbf{a}_{tp}^{{\rm H}}(\theta_{tp}^{11},\phi_{tp}^{11}),
	\end{equation}
	where $\mathbf{a}_{rp}(\theta_{rp}^{11},\phi_{rp}^{11})$ and $\mathbf{a}_{tp}(\theta_{tp}^{11},\phi_{tp}^{11})$ stand for the array response vectors of the $p^{\rm th}$ path at Tx and Rx, respectively. 
	Specifically, $\mathbf{a}_{tp}(\theta_{tp}^{11},\phi_{tp}^{11}) $ is expressed as
	\begin{equation}\label{equ_array_steering_vector}
	\mathbf{a}_{tp}(\theta_{tp}^{11},\phi_{tp}^{11})=\left[\!1\! \dots\! \mathrm{e}^{j\frac{2\pi d}{\lambda}(\psi_t-\psi_r)}\dots\!\ \mathrm{e}^{j\frac{2\pi d}{\lambda}(\Psi_t-\Psi_r)}\!\right]^{\mathrm{T}},
	\end{equation}
	where $\Psi_t = (M_{K_tx}+N^a_{K_tx}-2){\rm sin}\theta_{tp}^{11}{\rm cos}\phi_{tp}^{11}-(M_{K_tz}+N^a_{K_tz}-2){\rm sin}\phi_{tp}^{11}$, $N^a_{K_tx}$ and $N^a_{K_tz}$ denote the number of antennas of the $K_t^{\rm th}$ subarray along x-axis and z-axis, respectively. 
	Moreover, $\Psi_r = (M_{K_rx}+N^a_{K_rx}-2){\rm sin}\theta_{rp}^{11}{\rm cos}\phi_{rp}^{11}-(M_{K_rz}+N^a_{K_rz}-2){\rm sin}\phi_{rp}^{11}$, where $N^a_{K_rx}$ and $N^a_{K_rz}$ denote the number of antennas of the $K_r^{\rm th}$ subarray along x-axis and z-axis, respectively. 
	Similarly, $\mathbf{a}_{rp}(\theta_{rp}^{11},\phi_{rp}^{11})$ is constructed by replacing the superscript $t$ as $r$ in~\eqref{equ_array_steering_vector}.
	In this work, we deploy the ray-tracing method widely used to model the THz channels to obtain the deterministic multi-path channel, which does not account for the path clustering and angle distributions~\cite{ref_ray_tracing, DAoSA}.
		During the implementation, we deploy the narrowband communication system with bandwidth smaller than coherence bandwidth, which can be regarded as one sub-band of the THz multicarrier systems~\cite{ref_CE_GAN, ref_THz_Multicarrier}. Taking the narrowband system and limited array size of planar array into consideration, the beam squint effect in wideband THz systems contributes negligibly~\cite{ref_hybrid_magazine}.
	
	The parameter set $\mathcal{P}_{P} = \{|\alpha^{11}_p|, D^{11}_p, \theta^{11}_{tp},\phi^{11}_{tp}, \theta^{11}_{rp}, \phi^{11}_{rp} \}$ with $6N_p$ elements uniquely determines the planar-wave channel matrix in~\eqref{equ_P_model}. 
	Given the sparsity of THz channel that $N_p\leq10$~\cite{DAoSA}, the required number of parameters to determine the PWM is much smaller than the SWM in~\eqref{equ_SW_channel}.

	\subsection{Accuracy Analysis for the PWM in THz UM-MIMO Systems.}
	\label{subsec_ccuracy Analysis for the Planar-wave Channel Model in THz UM-MIMO Systems.}
	
	To evaluate the accuracy of the PWM, we consider the approximation error of the path gains for the LoS path, since which dominates the channel in the THz bands~\cite{ref_2level}, the analysis of other multi-path components is extensible.
	Particularly, we define the normalized approximation error of the path gain between the $l^{\rm th}$ transmitted antenna and the $i^{\rm th}$ received antenna as
	\begin{equation} \label{equ_error_planar_1}
	\epsilon^{il}=\frac{\overbrace{\left\lvert \lvert\alpha^{il}\rvert e^{-j\frac{2\pi}{\lambda}D^{il}}-\lvert\alpha^{11}\rvert e^{-j\frac{2\pi}{\lambda} D^{11}} e^{-j\frac{2\pi}{\lambda}\Delta D^{il}} \right\rvert }^{(*)}}{\lvert\alpha^{il}\rvert},
	\end{equation}
	where the first and second terms of $(*)$ are the path gains calculated by the SWM and PWM~\eqref{equ_gain} and~\eqref{equ_approx_gain}, respectively.
	Since the amplitude of the path gain $|\alpha^{il}|$ is proportional to $D^{il}$, which is treated as the same with different $i$ and $l$ by considering the array size is far less than the communication distance.
	We obtain $\lvert\alpha^{11}\rvert\approx\dots\approx\lvert\alpha^{N_rN_t}\rvert$, and~\eqref{equ_error_planar_1} is rewritten as 
	\begin{equation}
	\begin{split}
	\epsilon_{il}&\approx\left\lvert e^{- j\frac{2\pi}{\lambda}D^{il}}- e^{-j\frac{2\pi}{\lambda} D^{11}} e^{-j\frac{2\pi}{\lambda}\Delta D^{il}} \right\rvert.
	\end{split}
	\label{equ_error_planar_2}
	\end{equation}
	
	From~APPENDIX~\ref{Sec_Appendix}, we can further calculate $\epsilon_{il}$ as
	\begin{equation}\label{equ_approximation_error}
	\begin{split}
	\epsilon_{il}& \approx \Big|{\rm sin}\Big(\frac{\pi d^2}{2D^{11}\lambda} \left[(m_{k_rx}-m_{k_tx} + n_{k_rx} - n_{k_tx})^2
	\left({\rm sin}^2\theta^{11}_{t1} {\rm cos}^2\phi^{11}_{t1}\! + \!{\rm cos}^2\theta^{11}_{t1}\right)\right. \\
	& \left.+ (m_{k_rz}-m_{k_tz} + n_{k_rz} - n_{k_tz})^2{\rm cos}^2\phi^{11}_{t1}
	\right] + \mathcal{P}\Big)\Big|.
	\end{split}
	\end{equation}
	Therefore, $\epsilon_{il}$ under the PWM is positively related to $(m_{k_rx}-m_{k_tx} + n_{k_rx} - n_{k_tx})^2$ and $(m_{k_rz}-m_{k_tz} + n_{k_rz} - n_{k_tz})^2$, whereas negatively related to the communication distance $D^{11}$ and the carrier wavelength $\lambda$. 
	The maximum values of $(m_{k_rx}-m_{k_tx} + n_{k_rx} - n_{k_tx})^2$ and $(m_{k_rz}-m_{k_tz} + n_{k_rz} - n_{k_tz})^2$ are ${\rm max} \{|M_{K_tx}+N_{K_tx}|^2,|M_{K_rx}+N_{K_rx}|^2\} $ and ${\rm max} \{|M_{K_tz}+N_{K_tz}|^2,|M_{K_rz}+N_{K_rz}|^2\} $, respectively. 
	We define the unit of array size as $\mathcal{L}_t = \sqrt{(M_{K_tx}+N_{K_tx})^2 + (M_{K_tz}+N_{K_tz})^2}$, $\mathcal{L}_r = \sqrt{(M_{K_rx}+N_{K_rx})^2 + (M_{K_rz}+N_{K_rz})^2}$, respectively.
	Therefore, $d\mathcal{L}_t$ and $d\mathcal{L}_t$ denote the array apertures at Tx and Rx, respectively. 
	The value of $\frac{\pi d^2 \mathcal{L}_t\mathcal{L}_r}{\lambda D^{11}}$ defines the array far-field and near-field, which is similar to the condition defined by the Rayleigh distance $D_{ray} = \frac{2S^2}{\lambda}$ in~\cite{ref_WSMS}, where $S$ denotes the array aperture. 
	Particularly, by dividing $D_{ray}$ to both sides, the expression of Rayleigh distance can be written as $\frac{2S^2}{\lambda D_{ray}} = 1$.
	The only difference between $\frac{\pi d^2 \mathcal{L}_t\mathcal{L}_r}{\lambda D^{11}}$ and $\frac{2S^2}{\lambda D_{ray}}$ is one constant.
	Moreover, $\frac{\pi d^2 \mathcal{L}_t\mathcal{L}_r}{\lambda D^{11}}$ is also related to the Fraunhofer distance in~\cite{ref_Franh} when the antenna array is considered.
	Therefore, when $\frac{\pi d^2 \mathcal{L}_t\mathcal{L}_r}{\lambda D^{11}}$ is comparable to~$1$ as the
	array aperture increases and the communication distance decreases, the far-field condition is violated and the PWM becomes inaccurate.
	By contrast, when$\frac{\pi d^2 \mathcal{L}_t\mathcal{L}_r}{\lambda D^{11}}$ is far less than 1, the PWM remains accurate. 

	At micro-wave and mmWave frequencies, due to the relatively long wavelength and communication distance, the PWM is usually accurate.
	However, in THz UM-MIMO systems, the PWM becomes inaccurate owning to the significantly smaller wavelength and shorter communication distance. 
	For example, when carrier frequency $f = 3$~GHz and $D^{11} = 500$~m, and both Tx and Rx are equipped with 16 antennas, $\frac{\pi d^2 \mathcal{L}_t\mathcal{L}_r}{\lambda D^{11}}\approx 4.24\times 10^{-4}$, which is far less than 1. By contrast, with a comparable array size, when $f = 0.3$~THz and $D^{11} = 50$~m, $\frac{\pi d^2 \mathcal{L}_t\mathcal{L}_r}{\lambda D^{11}}\approx 0.424$. In this case, $\epsilon_{il}$ is non-negligible. 
	
	\subsection{Hybrid Spherical- and Planar-wave Channel Model}
	\label{subSec_Hybrid Spherical- and Planar-wave Channel Model}
	
	The approximation error for PWM is positively related to the array size according to~\eqref{equ_approximation_error}. Inspired by this, we investigate the HSPM for THz UM-MIMO systems with high accuracy. 
	Particularly, the PWM is employed within one subarray, which remains precise due to the relatively small array size. 
	Among the subarrays, the SWM is utilized to improve the modeling accuracy. 
	The sub-channel between the $k_{t}^{\rm th}$ transmitted subarray and the $k_{r}^{\rm th}$ received subarray is $\Sigma_{p=1}^{N_p}\alpha^{k_{r}k_{t}}_p\mathbf{a}_{rp}^{k_{r}k_{t}}(\mathbf{a}_{tp}^{k_{r}k_{t}})^{\rm H}$, where $\alpha^{k_{r}k_{t}}_p$ denotes the path gain between the reference antennas of the $k_t^{\rm th}$ transmitted and $k_r^{\rm th}$ received subarrays.
	Moreover, $\mathbf{a}_{tp}^{k_{r}k_{t}}=\mathbf{a}_{tp}^{k_rk_t}(\theta_{tp}^{k_rk_t},\phi_{tp}^{k_rk_t})$ and $\mathbf{a}_{rp}^{k_{r}k_{t}}=\mathbf{a}_{rp}^{k_rk_t}(\theta_{rp}^{k_rk_t},\phi_{rp}^{k_rk_t})$ represent the array response vector of the transmitted and received subarrays as~\eqref{equ_array_steering_vector}, respectively.
	Therefore, the HSPM for UM-MIMO systems boils down as
		\begin{equation}\label{equ_HSPM_model}
		\mathbf{H}_{\rm HSPM}=\sum_{p=1}^{N_p}\!\!\left[\!\!\!\begin{array}{ccc}
		\vert\alpha^{11}_p\vert e^{-j\frac{2\pi}{\lambda}D^{11}_p}\mathbf{a}_{rp}^{11} (\mathbf{a}_{tp}^{11})^{\rm H}&\ldots&	\vert\alpha^{11}_p\vert e^{-j\frac{2\pi}{\lambda}D^{1K_t}_p}\mathbf{a}_{rp}^{1K_t}(\mathbf{a}_{tp}^{1K_t})^{\rm H}\\
		\vdots&\ddots&\vdots\\
		\vert\alpha^{11}_p\vert e^{-j\frac{2\pi}{\lambda}D^{K_r1}_p}\mathbf{a}_{rp}^{K_r1}(\mathbf{a}_{tp}^{K_r1})^{\rm H}&\ldots&\vert\alpha^{11}_p\vert e^{-j\frac{2\pi}{\lambda}D^{K_rK_t}_p}\mathbf{a}_{rp}^{K_rK_t}(\mathbf{a}_{tp}^{\!K_rK_t})^{\rm H}\\
		\end{array}\!\!\!\!\right]. 
		\end{equation}
		In \eqref{equ_HSPM_model}, different subarrays share common reflectors, leading the number of multi-path is the same for different subarray pairs. 
		Moreover, the amplitude of path gain is the same for different subarrays, while the DoD, DoA and phase of path gain are different among different subarrays.
	In addition, the HSPM channel model is suitable for the THz band ranging from 0.1 to 10~THz.

	The HSPM in~\eqref{equ_HSPM_model} is characterized the parameter set $\mathcal{P}_{HSPM} =\left\{|\alpha_{p}^{11}|,D^{k_rk_t}_p, \theta_{rp}^{k_rk_t},\phi_{rp}^{k_rk_t},\right.$ $\left.\theta_{tp}^{k_rk_t},\phi_{tp}^{k_rk_t} \right\}$, which contains $N_p(1+5K_rK_t)$ elements.
	The spherical-wave UM-MIMO channel model~\eqref{equ_SW_channel} and the planar-wave UM-MIMO channel model~\eqref{equ_P_model} are special cases of the HSPM when $K_t=N_t, K_r=N_r$ and $K_t=K_r=1$, respectively. The high accuracy of the HSPM is evaluated in Sec.~\ref{subsec_Accuracy of HSPM Channel model}.
	Therefore, with various $K_t$ and $K_r$, the HSPM achieves high accuracy with relatively small number of channel parameters, compared to the PWM and SWM, respectively.

	\section{Two-phase Channel Estimation} \label{sec_Two-phase Channel Estimation}
	Existing CE algorithms are proposed for PWM and without considering the spherical-wave propagation, directly using which for HSPM CE incurs a huge performance loss~\cite{DCNN_GC}. 
	Moreover, $\mathbf{H}_{\textrm{HSPM}}$ in~\eqref{equ_HSPM_model} is hard to estimate, since which is of high dimensional for the typical THz UM-MIMO systems, e.g., $1024\times 1024$.
	Fortunately, only a few paths are available in THz channels, namely, $N_p \leq 10$. Performing parameters estimation processes inherent advantages of low complexity, as $N_p(1+5K_tK_r)$ is much smaller than the dimension of the channel matrix.
	Nevertheless, the value of $N_p(1+5K_tK_r)$ is still large when the number of subarrays increases. When $K_t=K_r=4$ and $N_p = 4$~\cite{DAoSA}, the required number of channel parameters to be estimated is $1296$. 
	To further reduce CE complexity, in this section, we propose a two-phase CE method, which first deploys a DCNN to directly estimate the channel parameters of the reference subarrays. 
	The remaining channel parameters among other subarrays are derived by exploiting the geometric relations. 
	Therefore, the required number of parameters to be estimated reduces to $6N_p$, which equals to the number of parameters in the PWM, thus leads to low complexity. 
	\subsection{Phase 1: DCNN for Channel Parameter Estimation}\label{subsec_DCNN for Channel Parameter Estimation}
	
	\subsubsection{Network Structure}
	\begin{figure}[t]
		\centering
		{\includegraphics[width=4in]{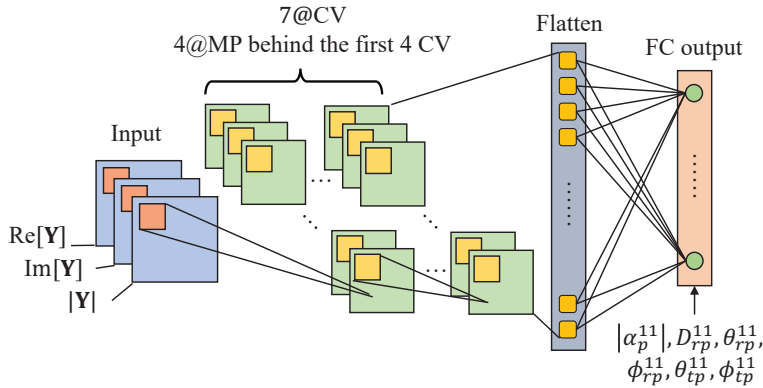}}
		\caption{The structure of the proposed DCNN network.}
		\label{Network_structure}
		\vspace{-6mm}
	\end{figure}
	The structure of the proposed DCNN network is illustrated in Fig.~\ref{Network_structure}, which estimates channel parameters between $T_1$ and $R_1$ in light of the channel observation in~\eqref{equ_Channel_observation}.
	In total, the DCNN network contains fifteen layers, including one input layer, seven convolutional (CV) layers, four max-pooling (MP) layers, one flattening layer and one fully-connected (FC) output layer. 
	The input layer is composed of three real-valued matrices obtained from the channel observation matrix in, including the element-wise real and imaginary values of the channel observation matrix $\mathbf{Y}$, denoted by ${\rm Re}[\mathbf{Y}]$ and ${\rm Im}[\mathbf{Y}]$, and the element-wise absolute value of $\mathbf{Y}$, expressed as $|\mathbf{Y}|$, respectively. 
	${\rm Re}[\mathbf{Y}]$, ${\rm Im}[\mathbf{Y}]$ fully describe the channel information, while $|\mathbf{Y}|$ elaborates the influence of the amplitude of path gain and the effect of noise.
	
	Followed by the input layer, there are seven CV layers, four MP layers, and one flatting layer. 
	The CV layers contain 16, 32, 64, 128, 62, 32, 16 filters, respectively, in which a convolution filter with size $3\times3$ is deployed to extract the features of the network input. 
	In the CV layer, each neuron connects to a small region of the previous layer called the local receptive field. 
	The convolution operation is performed to the $i^{\textrm{th}}$ local receptive field in the $(m-1)^{\textrm{th}}$ layer $\mathbf{C}_{\textrm{cv},i}^{(m-1)}$ by the convolutional filter.
	By denoting the weight and bias for the $j^{\textrm{th}}$ filter in $m^{\textrm{th}}$ layer as $\mathbf{u}_{\textrm{cv},j}^{(m)}$ and ${b}_{\textrm{cv},j}^{(m)}$, respectively, the value of the $i^{\textrm{th}}$ output neuron ${z}_{\textrm{cv},i}^{(m)}$ in CV is calculated as 
	\begin{equation}
	{z}_{\textrm{cv},i}^{(m)}=
	f^{(m)}\left(\mathbf{u}_{\textrm{cv},j}^{(m)}*\mathbf{C}_{\textrm{cv},i}^{({m-1})}+{b}_{\textrm{cv},j}^{(m)}\right),
	\end{equation} 
	where $f^{(m)}\left(\cdot\right)$ refers to the activation function describing the non-linear mapping relationship. 
	In addition, zero-padding (ZP) and batch-normalization (BN) are invoked during the convolution process. ZP is useful to maintain the dimension of CV layers by adding zeros in the marginal local receptive fields, while BN avoids possible gradient dispersion and speeds up the training process with normalization in the mini-batches. Moreover, the MP layers with a pool size of $2\times 2$ are inserted behind the first 4 CV layers, in which the maximum value in the pool is extracted to reduce the network dimension and simplify the training process.
	
	After the last CV layer, a flatting layer rearranges the neurons into one dimension and connects to the FC output layer. The neurons between neighboring layers in FC are fully-connected. In this way, given the output of the $(m-1)^{\textrm{th}}$ layer $\mathbf{z}^{({m-1})}_{\textrm{fc}}$, the value of the $i^{\textrm{th}}$ output neuron at the $m^{\textrm{th}}$ layer of a FC ${z}_{\textrm{fc},i}^{(m)}$ can be expressed as
	\begin{equation}\label{FC_output}
	{z}_{\textrm{fc},i}^{(m)}=f^{(m)}\left(\mathbf{u}_{\textrm{fc},i}^{(m)T}\mathbf{z}_{\textrm{fc}}^{({m-1})}+{b}_{\textrm{fc},i}^{(m)}\right),
	\end{equation}
	where $\mathbf{u}_{\textrm{fc},i}^{(m)}$ and ${b}_{\textrm{fc},i}^{(m)}$ are the weight and bias, respectively. 
	In the input layer and the hidden layers, the rectified linear unit (ReLU) $f_{\text{ReLU}}(x)={\rm max}\{0,x\}$ is explored for its fast computation speed.
	We choose the sigmoid function $f_{\text{Sigmoid}}(x) = \frac{1}{1+e^{-x}}$ at the output layer,
	which stables the output of the DCNN by restricting output range into $[0,1]$. 
	The connection of these aforementioned layers completes the DCNN network. By denoting the total number of layers as $M$ and the input as $\mathbf{x}$, respectively, the output $\mathbf{z}^{(M)}$ of the proposed network is represented as
	\begin{equation}
	\mathbf{z}^{(M)}=f^{(M)}\left(f^{({M-1)}}\left(\cdots f^{(1)}\left(\mathbf{x}\right)\right)\right),
	\end{equation}
	where $f^{(M)}$ refers to the activation function of the $M^{\rm th}$ layer.

	\subsubsection{Operation Policy}
	The parameters $\{|\alpha_{p}^{11}|,D^{11}_p, \theta_{rp}^{11},\phi_{rp}^{11},\theta_{tp}^{11},\phi_{tp}^{11} \}$ are chosen as training labels, which have various ranges. The amplitude of channel gain $|\alpha_{p}^{11}| \in [0,1]$, the communication distance $D^{11}_p\in [0,\infty]$, and the angles $\theta_{rp}^{11},\phi_{rp}^{11},\theta_{tp}^{11},\phi_{tp}^{11} \in [0,2\pi]$. 
	To facilitate the convergence of the network, both network input and training label are normalized into [0, 1].
We select the min-max normalization, since it achieves smaller training loss than the Sigmoid counterpart, as shown in Fig.~3. Furthermore, compared to z-score normalization that follows the normal distribution, both network input and training label in our network are normalized within [0,1]~\cite{ref_normalization}.
	By denoting the value to be normalized as $\chi$ and the minimum and maximum values of $\chi$ as $\chi_{\rm{min}}$ 
	and $\chi_{\rm{max}}$, respectively, the min-max normalization operates as $
	\tilde{\chi}=\left(\chi-\chi_{\rm{min}}\right)/\left(\chi_{\rm{max}}-\chi_{\rm{min}}\right).$
	\begin{figure}[t]
		\centering 
		\includegraphics[width=0.5\textwidth]{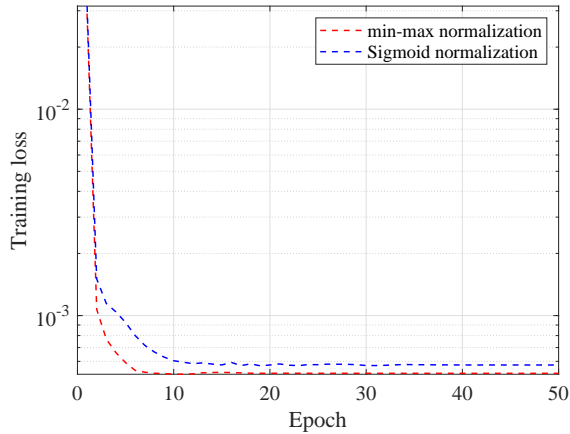}
		\caption{Training loss with different normalization methods.}
		\vspace{-6mm}
		\label{fig_norm_comp}
	\end{figure}

	To minimize the estimation error, we consider three losses that account for the accuracy of the angles $L_{\rm{angle}}=\mathbb{E} 
	\{\Arrowvert{\boldsymbol{\varsigma}}-\hat{\boldsymbol{\varsigma}}\Arrowvert/ \Arrowvert{\boldsymbol{\varsigma}}\Arrowvert \}$, distance $L_{\rm{dist}} =\mathbb{E}\{\Arrowvert\hat{D_{p}^{11}}-D_p^{11}\Arrowvert/\Arrowvert D_p^{11}\Arrowvert\}$ and amplitude of path gain $L_{\rm{gain}}=\mathbb{E}\{\Arrowvert\hat{|\alpha_{p}^{11}|}-|\alpha_{p}^{11}|\Arrowvert/\Arrowvert|\alpha_{p}^{11}|\Arrowvert\}$, where ${\boldsymbol{\varsigma}} = [\theta_{rp}^{11},\phi_{rp}^{11},\theta_{tp}^{11},\phi_{tp}^{11} ]$ and $\hat{\boldsymbol{\varsigma}}$ represent the true and estimated angle vectors, respectively.
	The loss function $L_{\rm{loss}}$ is represented as
	\begin{equation}\label{Loss_func}
	L_{\rm{loss}}=\iota_1L_{\rm{angle}}+\iota_2L_{\rm{dist}}+\iota_3L_{\rm{gain}},
	\end{equation}
	where $\iota_1$, $\iota_2$ and $\iota_3$ are coefficients to account for different weights. Moreover, the adaptive moment estimation (Adam) is selected as the optimizer owning to its fastest convergence rate.
	
	\subsection{Phase 2: Parameters Extension via Geometric Relationships}\label{subsec_HSPM Channel Estimation}
	
\begin{figure}[t]
		\setlength{\belowcaptionskip}{0pt}
		\centering
		\subfigure[x-y plane.]{
			\includegraphics[width=0.33\textwidth]{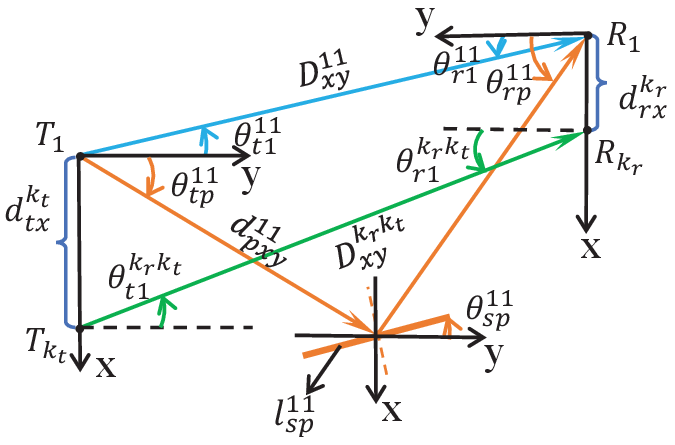} \label{fig_geometric_xy}}
		\hspace{0.6in}
		\subfigure[y-z plane.]{
			\includegraphics[width=0.33\textwidth]{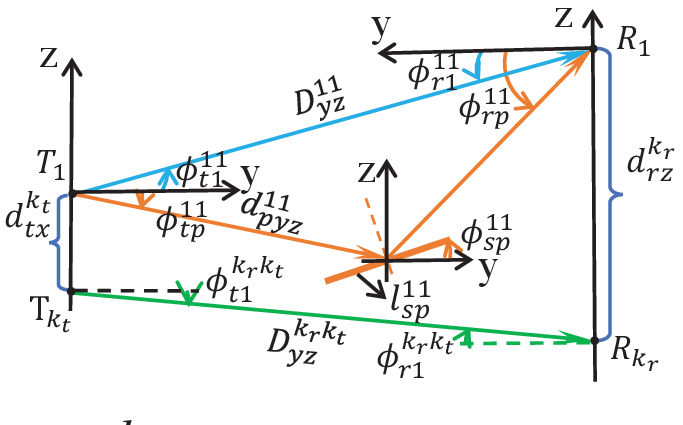} \label{fig_geometric_yz}}
		\caption{Projection of the communication paths.}
		\label{fig_geometric}
		\vspace{-6mm}
	\end{figure}

	After estimating the parameters between $T_1$ and $R_1$ through the DCNN,
	we use the geometric relationship to derive the channel parameters of other subarrays. 
	Specifically, 
	we derive the channel parameters between the $k_t^{\rm th}$ subarray at Tx $T_{k_t}$ and the $k_r^{\rm th}$ subarray at Rx $R_{k_r}$, respectively, including the DoD, DoA pairs $(\theta_{tp}^{k_tk_r}, \phi_{tp}^{k_tk_r})$, $(\theta_{rp}^{k_tk_r}, \phi_{rp}^{k_tk_r})$ and the communication distance $D_{p}^{k_tk_r}$.
	As illustrated in Fig.~\ref{fig_system_model}, $S_p^{11}$ stands for the reflection point between the reference antennas at Tx and Rx. 
	At $S^{11}_p$, $l_{sp}^{11}$ refers to the intersection line between the reflector and the plane of incoming and reflected rays. 
	The azimuth and elevation angles of $l_{sp}^{11}$ are denoted as $\theta^{11}_{sp}$ and $\phi^{11}_{sp}$, respectively.
	Moreover, $d_p^{11}$ and $D_p^{11}$ refers to the distance between the reference antenna at Tx and the reflector, as well as the communication distance of the $p^{\rm th}$ path, respectively. 
	By projecting the propagation paths in Fig.~\ref{fig_system_model} to the x-y and y-z plane, respectively, we obtain Fig.~\ref{fig_geometric}, where the
	definitions of $D^{11}_{xy}$, $D^{k_rk_t}_{xy}$, $d^{11}_{pxy}$, $D^{11}_{yz}$, $D^{k_rk_t}_{yz}$ and $d^{11}_{pyz}$ are stated in the following.
	\subsubsection{Equation of the Reflector}
	We first derive the equation of the reflector as preliminary result. 
	As illustrated in Fig.~\ref{fig_geometric}, $\theta^{11}_{sp}$ and $\phi^{11}_{sp}$ are calculated as
	$\theta^{11}_{sp} = \frac{1}{2}\left(\theta^{11}_{tp}-\theta_{rp}^{11}\right)$ and $ \phi^{11}_{sp} = \frac{1}{2}\left(\phi^{11}_{tp}-\phi_{rp}^{11}\right)$, respectively. We have
	$
	\frac{{\rm sin}(\theta_{rp}^{11}-\theta_{r1}^{11} )}{d^{11}_{pxy}}=
	\frac{{\rm sin}(180^\circ-\theta^{11}_{rp} - \theta^{11}_{tp})}{D^{11}_{xy}},
	$ where $d^{11}_{pxy} = d^{11}_{p} { \rm cos}\phi_{tp}^{11}$, $D^{11}_{xy} = D^{11} {\rm cos}\phi_{t1}^{11}$ and
	$
	d^{11}_{p} =
	\frac{D^{11}_{xy}{\rm sin}(\theta_{rp}^{11}-\theta_{r1}^{11} )}{{\rm sin}(\theta^{11}_{rp} + \theta^{11}_{tp})\rm{cos}\phi_{tp}^{11}}.$
	We consider the coordinate system at Tx, and set $T_{1} = (0,0,0)$, we have
	$S_p^{11} \!= \!(d^{11}_{p}{\rm sin}\theta_{tp}^{11}{\rm cos}\phi_{tp}^{11}\!, \!d^{11}_{p}{\rm cos}\theta_{tp}^{11}{\rm cos}\phi_{tp}^{11}\!,\!d^{11}_{p}{\rm sin}\phi_{tp}^{11})$ $= (S_{px}^{11},S_{py}^{11}, S_{pz}^{11})$, $R_{1} = (D^{11}{\rm sin}\theta_{t1}^{11}{\rm cos}\phi_{t1}^{11},$ $ D^{11}{\rm cos}\theta_{t1}^{11}{\rm cos}\phi_{t1}^{11},$ $D^{11}{\rm sin}\phi_{t1}^{11}) = (R_{1x},R_{1y},R_{1z})$. 
	These points uniquely determine the plane of incoming and reflected rays with equation $A_p^{11}x + B_p^{11}y +C_p^{11}z +D_p^{11}=0$. Since $T_1$, $S_p^{11}$ and $R_{1}$ are on this plane, we obtain $D_p^{11} = 0$ and the following equations as
	\begin{equation}
	\left\{ \begin{aligned}
	&A_p^{11}{\rm sin}\theta_{tp}^{11}{\rm cos}\phi_{tp}^{11} + B_p^{11}{\rm cos}\theta_{tp}^{11}{\rm cos}\phi_{tp}^{11} +C_p^{11}{\rm sin}\phi_{tp}^{11}=\!0, \\ 
	&
	A_p^{11}{\rm sin}\theta_{t1}^{11}{\rm cos}\phi_{t1}^{11} + B_p^{11}{\rm cos}\theta_{t1}^{11}{\rm cos}\phi_{t1}^{11} +C_p^{11}{\rm sin}\phi_{t1}^{11}=0.
	\end{aligned} 
	\right. 
	\end{equation}
	By setting $C_p^{11}=1$, we obtain one of the solutions as
	\begin{equation}\label{equ_planeinputoutput}
	\left\{ 
	\begin{aligned}
	&\!\!A_p^{11}\!=\!\frac{{\rm cos}\theta_{tp}^{11}{\rm cos}\phi_{tp}^{11}{\rm sin}\phi_{t1}^{11} -{\rm cos}\theta_{t1}^{11} {\rm cos}\phi_{t1}^{11}{\rm sin}\phi_{tp}^{11} }{{\rm cos}\phi_{tp}^{11} {\rm cos}\phi_{t1}^{11} \left({\rm cos}\theta_{tp}^{11}{\rm sin}\theta_{t1}^{11} - {\rm cos}\theta_{t1}^{11}{\rm sin}\theta_{tp}^{11}\right)},\\
	&\!\!	B_p^{11}\!=\!\frac{ {\rm cos}\phi_{tp}^{11}{\rm sin}\phi_{t1}^{11}{\rm sin}\theta_{tp}^{11}-{\rm cos}\phi_{t1}^{11}{\rm sin}\phi_{tp}^{11}{\rm sin}\theta_{t1}^{11} }{{\rm cos}\phi_{t1}^{11} {\rm cos}\phi_{tp}^{11} \left({\rm cos}\theta_{tp}^{11}{\rm sin}\theta_{t1}^{11}- {\rm cos}\theta_{t1}^{11}{\rm sin}\theta_{tp}^{11}\right)}.
	\end{aligned} 
	\right. 
	\end{equation}
	Since $({\rm sin}\theta_{sp}^{11}{\rm cos}\phi_{sp}^{11}, {\rm cos}\theta_{sp}^{11}{\rm cos}\phi_{sp}^{11},{\rm sin}\phi_{sp}^{11})$ indicates the direction of $l^{11}_{sp}$, there are
	$\frac{x-d^{11}_{p}{\rm sin} \theta_{tp}^{11}{\rm cos}\phi_{tp}^{11}}{{\rm sin}\theta_{sp}^{11}\!{\rm cos}\phi_{sp}^{11}}
	$ $=\frac{y- d^{11}_{p}{\rm cos}\theta_{tp}^{11}{\rm cos}\phi_{tp}^{11}}{{\rm cos}\theta_{sp}^{11}{\rm cos}\phi_{sp}^{11}} =\frac{z-d^{11}_{p}{\rm sin}\phi_{tp}^{11}}{{\rm sin}\phi_{sp}^{11}}.
	$
	By setting $z = 0$, we can find $S_{p0}^{11}$ $= \bigg (d^{11}_{p}\Big({\rm sin}\theta_{tp}^{11}{\rm cos}\phi_{tp}^{11}\!-\!\frac{{\rm sin}\theta_{sp}^{11}{\rm cos}\phi_{sp}^{11}{\rm sin}\phi_{tp}^{11}}{{\rm sin}\phi_{sp}^{11}}\Big), d^{11}_{p}\Big({\rm cos}\theta_{tp}^{11}{\rm cos}\phi_{tp}^{11}\!-\!\frac{{\rm cos}\theta_{sp}^{11}{\rm cos}\phi_{sp}^{11}{\rm sin}\phi_{tp}^{11}}{{\rm sin}\phi_{sp}^{11}}\Big), 0 \bigg)$ on $l^{11}_{sp}$.
	Consider the plane equation of the $p^{\rm th}$ reflector as $A_px + B_py +C_pz +D_p=0$, therefore, there are
	\begin{equation}\label{reflector2}
	\left\{ 
	\begin{aligned}
	&\!\!A_pd^{11}_{p}{\rm sin}\theta_{tp}^{11}{\rm cos}\phi_{tp}^{11} + B_pd^{11}_{p}{\rm cos}\theta_{tp}^{11}{\rm cos}\phi_{tp}^{11} +C_pd^{11}_{p}{\rm sin}\phi_{tp}^{11} +D_p=0,\\
	&\!\!A_p\Big({\rm sin}\theta_{tp}^{11}{\rm cos}\phi_{tp}^{11}\!\!-\!\!\frac{{\rm sin}\theta_{sp}^{11}{\rm cos}\phi_{sp}^{11}{\rm sin}\phi_{tp}^{11}}{{\rm sin}\phi_{sp}^{11}}\Big) \!+\!B_p \!\Big(\!{\rm cos}\theta_{tp}^{11}{\rm cos}\phi_{tp}^{11}\!\!-\!\!\frac{{\rm cos}\theta_{sp}^{11}{\rm cos}\phi_{sp}^{11}{\rm sin}\phi_{tp}^{11}}{{\rm sin}\phi_{sp}^{11}}\!\Big) \!\!+\!\! \frac{D_p}{d^{11}_{p}}\!\!=\!\!0,\\
	&A_pA_p^{11} + B_pB_p^{11}+ C_pC_p^{11} = 0.
	\end{aligned} 
	\right.
	\end{equation}
	By choosing $C_p=1$, one of the solutions is calculated as
	\begin{equation} \label{equ_27}
	\left\{ 
	\begin{aligned}
	&A_p=\frac{{\rm sin}\phi_{tp}^{11}\left( C_p^{11}
		{\rm cos}\theta_{sp}^{11}{\rm cos}\phi_{sp}^{11} -B_p^{11}{\rm sin}\phi_{tp}^{11}\right)
	}{{\rm cos}\phi_{sp}^{11}{\rm sin}\phi_{tp}^{11}\left(A_p^{11}{\rm cos}\theta_{sp}^{11} - B_p^{11}{\rm sin}\theta_{sp}^{11}\right)
	},\\
	&B_p=\frac{{\rm sin}\phi_{tp}^{11}
		\left( A_p^{11}{\rm sin}\phi_{sp}^{11} -C_p^{11}{\rm sin}\theta_{sp}^{11}{\rm cos}\phi_{sp}^{11}\right)}{
		{\rm cos}\phi_{sp}^{11}{\rm sin}\phi_{tp}^{11}\left(A_p^{11}{\rm cos}\theta_{sp}^{11} - B_p^{11}{\rm sin}\theta_{sp}^{11}\right)}. 
	\end{aligned} 
	\right. 
	\end{equation}
	Therefore, we can obtain the equation of the reflector. 
	
	\subsubsection{Channel Parameters for LoS path}\label{subsubsec_Channel Parameters for LoS path}
	The channel parameters for the LoS path are calculated as follows. 
	In Fig~\ref{fig_geometric_xy}, we have $
	D^{k_tk_r}_{xy}
	=\sqrt{ \!(\Delta d_{x}^{k_tk_r})^2\!+\!(D^{11}_{xy})^2\!-\!2\Delta d_{x}^{k_tk_r}D^{11}_{xy}{\rm sin}\theta_{t1}^{11} },
	$
	where $D^{k_tk_r}_{xy} \!\!=\!\! D^{k_tk_r}{\rm cos}{\phi_{tp}^{k_tk_r}}$, $\Delta d_{x}^{k_tk_r}\! \!= \!\!d_{tx}^{k_t}-d_{rx}^{k_r}$.
	And
	$\frac{{\rm sin}(90^\circ +\theta^{k_tk_r}_{t1})}{D^{11}_{xy}}=	\frac{{\rm sin}(\theta^{k_tk_r}_{r1} -\theta^{11}_{r1})}{\Delta d_x^{k_tk_r}}=
	\frac{{\rm sin}(90^\circ-\theta^{11}_{t1})}{D^{k_tk_r}_{xy}}.$
	Similarly, in the y-z plane, we have
	$
	D^{k_tk_r}_{yz}=\sqrt{(\Delta d_{z}^{k_tk_r})^2+(D^{11}_{yz})^2-2\Delta d_{z}^{k_tk_r}D^{11}_{yz}{\rm sin}\phi_r^{11} },
	$
	where $D^{k_tk_r}_{yz} = D^{k_tk_r}{\rm cos}{\theta_t^{k_tk_r}}$, $\Delta d_z^{k_tk_r} = d_{rz}^{k_t}-d_{tz}^{k_r}$, $D^{11}_{yz} =D^{11}{\rm cos}\theta_{t}^{11} $.
	And
	$
	\frac{{\rm sin}(90^\circ +\phi^{k_tk_r}_{t})}{D^{11}_{yz}}=	\frac{{\rm sin}(\phi^{11}_{t} + \phi^{k_tk_r}_{r})}{\Delta d_z^{k_tk_r}}=
	\frac{{\rm sin}(90^\circ-\phi^{11}_{t})}{D^{k_tk_r}_{yz}}.
	$
	Therefore, the LoS angles and distance can be calculated as
	\begin{subequations}\label{equ_los_para}
		\begin{align}
		\theta_{t}^{k_tk_r} &= {\rm arccos}\left(\!\frac{D^{11}_{xy}{\rm cos}\theta_t^{11}}{D^{k_tk_r}_{xy}}\!\right),\\
		\theta_{r}^{k_tk_r}&= \theta_{r}^{11}+{\rm arcsin}{\left(\!\frac{\Delta d_x {\rm cos}\theta_t^{11}}{D^{k_tk_r}_{xy}}\!\right)},\\
		\phi_{t}^{k_tk_r} &={\rm arccos}{\left(\frac{D^{11}_{yz}{\rm cos}\phi_{t1}^{11}}{D^{k_tk_r}_{yz}}\right)},
		\\
		\phi_{r}^{k_tk_r} &= -\phi_{r}^{11}+{\rm arcsin}{\left(\frac{\Delta d_z{\rm cos}\phi_{t1}^{11}}{D^{k_tk_r}_{yz}}\right)},\\
		D^{k_tk_r} &=\frac{D^{k_tk_r}_{yz}}{{\rm cos}{\theta_{t1}^{k_tk_r}}}=\frac{D^{k_tk_r}_{xy}}{{\rm cos}{\phi_{t1}^{k_tk_r}}}.
		\end{align}
	\end{subequations} 
	
	\subsubsection{Channel Parameters for NLoS paths}\label{subsubsec_Channel Parameters for NLoS paths}
	The channel parameters for the NLoS paths are calculated as follows.
	We have $T_{k_t} =(d_{tx}^{k_t}, 0,-d_{tz}^{k_t})$ and $R_{k_r} =(D^{11}{\rm sin}\theta_{t1}^{11}{\rm cos}\phi_{t1}^{11}+d_{rx}^{k_r}, D^{11}{\rm cos}\theta_{t1}^{11}{\rm cos}\phi_{t1}^{11}, $ $
	D^{11}{\rm sin}\phi_{t1}^{11}-d_{rz}^{k_r}) = (R_{k_rx},R_{k_ry},R_{k_rz})$. 
	Suppose $S_p^{k_tk_r} = (x^{k_tk_r},y^{k_tk_r},z^{k_tk_r})$. $T_{k_t}$, $R_{k_r}$ and $S_p^{k_tk_r}$ determine the plane of the incoming and reflected rays with equation $A_p^{k_tk_r}x + B_p^{k_tk_r}y +C_p^{k_tk_r}z +D_p^{k_tk_r}=0$. Since $T_{k_t}$, $R_{k_r}$ and $S_p^{k_tk_r}$ are on this plane, we derive the following equations
	\begin{equation}
	\left\{ 
	\begin{aligned}
	&\!A_p^{k_tk_r}d_{tx}^{k_t} -C_p^{k_tk_r}d_{tz}^{k_t} +D_p^{k_tk_r}=0,\\
	&\!A_p^{k_tk_r}R_{k_rx} \!+\! B_p^{k_tk_r}R_{k_ry} +C_p^{k_tk_r}R_{k_rz} +D_p^{k_tk_r}=0,\\ 
	&\!A_p^{k_tk_r}x^{k_tk_r} + B_p^{k_tk_r}y^{k_tk_r} +C_p^{k_tk_r}z^{k_tk_r} +D_p^{k_tk_r}=0.
	\end{aligned} 
	\right. 
	\end{equation}
	By setting $C_k^{k_tk_r}=1$, we obtain one of the solutions for $A_p^{k_tk_r},B_p^{k_tk_r}$ as before.
	
	\begin{table}[t]
		\centering
		\renewcommand
		\arraystretch{} 
		\begin{tabular}{l}
			\toprule
			\textbf{Algorithm 1:} \textbf{DCNN for Channel Estimation} \\
			\midrule 
			\textbf{Input}: $\mathbf{Y}$\\
			1.~Obtain $\{|\alpha_{p}^{11}|,D^{11}_p, \theta_{rp}^{11},\phi_{rp}^{11},\theta_{tp}^{11},\phi_{tp}^{11} \}$ by DCNN.\\
			2.~\textbf{for} $k_t = 1,...,K_t-1$\\
			3.~~~~~\textbf{for} $k_r = 1,...,K_r-1$
			\\
			4.~~~~~~~Calculate LoS channel parameters by~\eqref{equ_los_para}. \\
			5.~~~~~~~Calculate NLoS channel parameters by~\eqref{equ_nlos_para}. \\
			6.~~~~~\textbf{end~for}\\
			7.~~~\textbf{end~for}\\
			8.~Reconstruct $\mathbf{H}$ in~\eqref{equ_P_model} \\
			\textbf{Output}: $\mathbf{H}$\\
			\bottomrule
		\end{tabular}
		\vspace{-6mm}
	\end{table}
	
	By the reflection theorem, the plane of the incoming and reflected rays is orthogonal to the reflector plane, and $S_p^{k_tk_r}$ is on the plane of the reflector, we obtain the following equations
	\begin{equation}\label{27}
	\left\{ 
	\begin{aligned}
	&\frac{	A_px^{k_tk_r} +B_py^{k_tk_r}+C_p z^{k_tk_r} }{\left|\left[\!A_p,B_p,C_p\!\right]
		\right|
		\left|\left[\begin{matrix}
		x^{k_tk_r},y^{k_tk_r},z^{k_tk_r}
		\end{matrix}\right]\right|} = \frac{	\!	A_p(\!x^{k_tk_r}\!\!-\!\!R_{k_rx}\!)\! +\! B_p(\!y^{k_tk_r}\!-\!R_{k_ry}\!)\!+\!C_p(\!z^{k_tk_r}\!-\!R_{k_rz}\!) \!}{|\left[A_p,B_p,C_p\right]|\left|	\left[\begin{matrix}
		x^{k_tk_r}-R_{k_rx}&
		\\y^{k_tk_r}-R_{k_ry}&
		\\z^{k_tk_r}-R_{k_rz}&
		\end{matrix}
		\!\!\!\!\!\right]\right|},\\
	&A_pA_p^{k_tk_r} + B_pB_p^{k_tk_r}+ C_pC_p^{k_tk_r} = 0,
	\\& A_p(x^{k_tk_r}\!-\!S_{px}^{11})\! +\! B_p(y^{k_tk_r}\!-\!S_{py}^{11}) \!+\!C_p(z^{k_tk_r}\!-\!S_{pz}^{11})=0, 
	\end{aligned} 
	\right. 
	\end{equation} 
	where the norms in equation~\eqref{27} are calculated as
	$|[A_p,B_p,C_p ]| = \sqrt{(A_p)^2+ (B_p)^2 + (C_p)^2}$, $
	|[x^{k_tk_r},y^{k_tk_r},z^{k_tk_r}]|
	=\sqrt{(x^{k_tk_r})^2+(\!y^{k_tk_r})^2 + (z^{k_tk_r})^2}$ 
	and 
	$|[x^{k_tk_r}-R_{k_rx},y^{k_tk_r}-R_{k_ry},z^{k_tk_r}-R_{k_rz}]^{\rm T}| $ $=\sqrt{\left[x^{k_tk_r}\!-\!R_{k_rx}\right]^2 \!+\! \left[y^{k_tk_r}\!-\!R_{k_ry}\!\right]^2 \!+\! \left[ z^{k_tk_r}\!-\!R_{k_rz}\!\right]^2\!}.
	$ We use the Newton method to solve the above equations to obtain $S_p^{k_tk_r}$, in which the initial value of $S_p^{k_tk_r}$ is set as $S_p^{11}$. As a result, the angles and distance for the $p^{\rm th} $ NLoS path are calculated as
	\begin{subequations}\label{equ_nlos_para}
		\begin{align}
		\theta^{k_tk_r}_{tp}&={\rm arcsin}\left[\!\frac{x^{k_rk_r}_{p}}{\sqrt{(x^{k_rk_r}_{p})^{2} + (y^{k_rk_r}_{p})^{2}}}\!\right],\\
		\phi^{k_tk_r}_{tp}&={\rm arcsin}\left[\!\frac{z^{k_rk_r}_{p}}{\sqrt{(x^{k_rk_r}_{p})^{2} + (y^{k_rk_r}_{p})^{2}+ (z^{k_rk_r}_{p})^{2}}}\!\right],\\
		\theta^{k_tk_r}_{rp}&={\rm arcsin}\left[\!\frac{x^{k_rk_r}_{p} -R_{k_rx}}{\sqrt{(x^{k_rk_r}_{p} \!-\!R_{1x})^{2} \!+\! (y^{k_rk_r}_{p}\!-\!R_{1y})^{2}}}\!\right],\\
		\phi^{k_tk_r}_{rp}&={\rm arcsin}\left[\!\frac{z^{k_rk_r}_{p}-R_{k_rz}}{\sqrt{\!(x^{k_rk_r}_{p} -R_{k_rx} )^{2} + (y^{k_rk_r}_{p}-R_{k_ry})^{2}+ (z^{k_rk_r}_{p}-R_{k_rz})^2}}\!\right],\\
		D_{p}^{k_tk_r}&= \sqrt{(x^{k_rk_r}_{p})^2 + (y^{k_rk_r}_{p})^2 + (z^{k_rk_r}_{p})^2}\\\notag&+ \sqrt{ (R_{k_rx}-x^{k_rk_r}_{p})^2 + (R_{k_ry}-y^{k_rk_r}_{p})^2 + (R_{k_rz}-z^{k_rk_r}_{p})^2}. 
		\end{align}
	\end{subequations}
	
	After obtaining all the channel parameters, the channel matrix in~\eqref{equ_HSPM_model} is recovered to complete the CE process. The proposed CE process is summarized in \textbf{Algorithm 1}.

	\section{Performance Evaluation}\label{sec_Performance_Evaluation}
	
	In this section, the accuracy of the proposed HSPM, and the performance of the proposed DCNN CE method are extensively evaluated. 
	\subsection{Evaluation Setup}

	\subsubsection{Simulation Environment}
\begin{figure}[t]
		\setlength{\belowcaptionskip}{0pt}
		\centering
		\subfigure[Simulation environment.]{
			\includegraphics[width=0.33\textwidth]{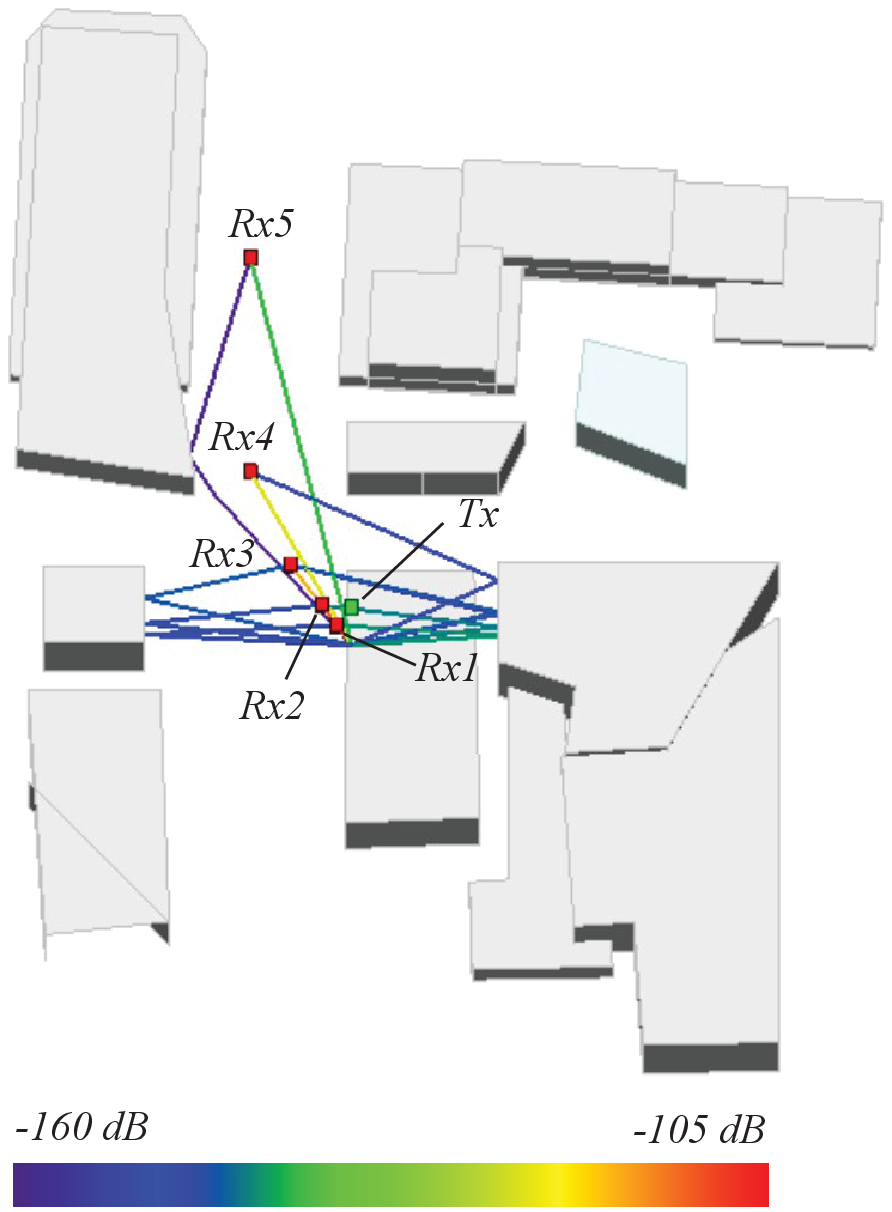} \label{fig_Simulation_environment}}
		\hspace{0.6in}
		\subfigure[ DoA and path gain of paths,  $D^{11}=80$~m, $f=0.8$~THz.]{
			\includegraphics[width=0.27\textwidth]{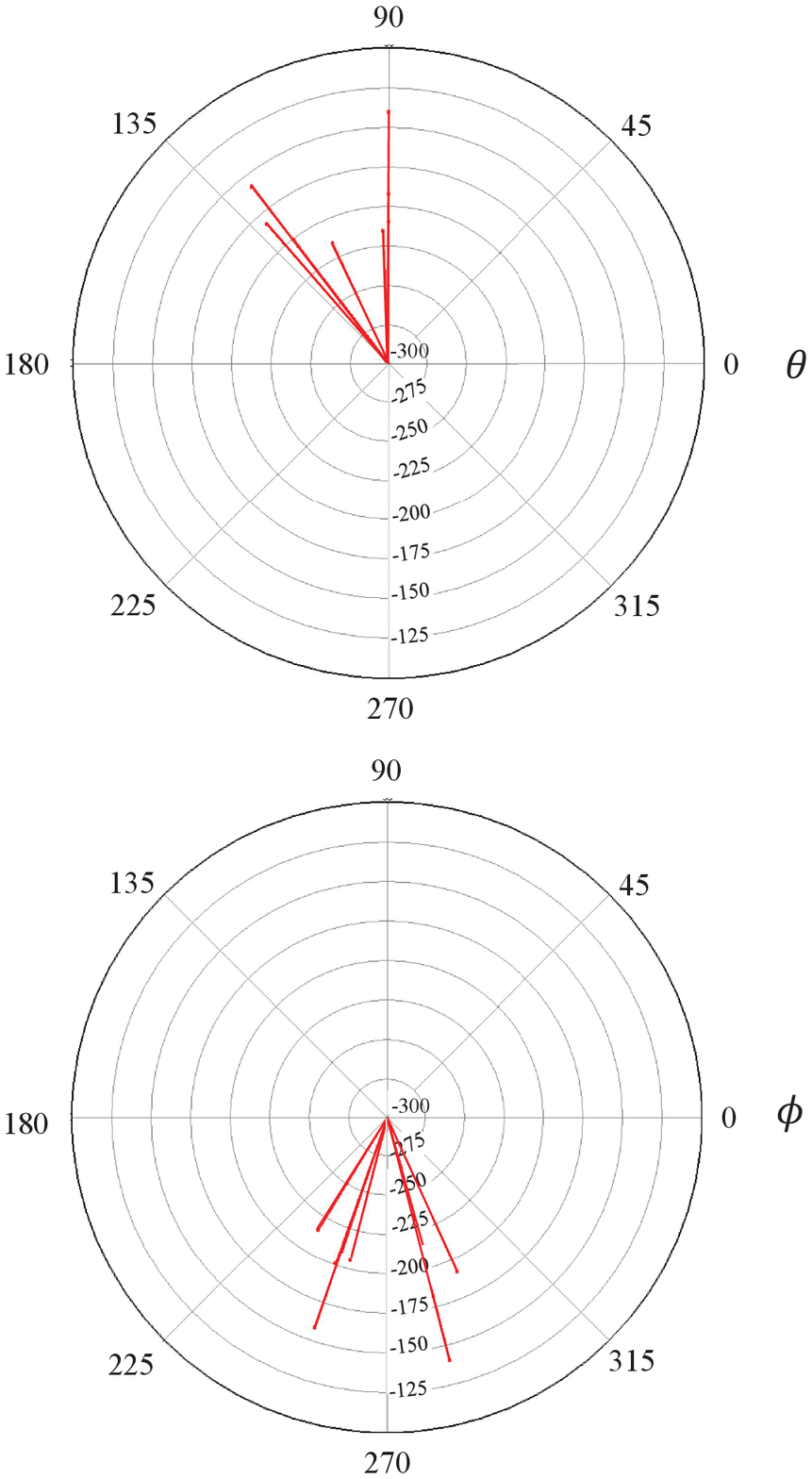} \label{fig_WI_angles}}
		\caption{The simulation environment and channel characteristics in the THz band.}
		\label{fig_WI}
		\vspace{-6mm}
	\end{figure}

	Due to the hardware constraint such as the difficulty in producing large arrays in the THz band, the real-life data in the THz UM-MIMO systems is currently unavailable.
		Therefore, we select the simulated THz channel to train the model.
		There are several simulators for simulating the THz channel, including NYUSIM~\cite{ref_NYUSIM}, TeraMIMO~\cite{ref_Tera_MIMO}, Wireless InSite~\cite{ref_wireless_insite} and etc. 
		In this work, we adopt Wireless InSite as suggested in~\cite{ref_wireless_insite}, which can well model the multi-path propagation based on ray-tracing techniques.
	As illustrated in Fig.~\ref{fig_Simulation_environment}, we consider a typical street scenario, in which several outdoor concrete buildings with different heights and flat terrain are considered~\cite{ref_urban}.
	We show an example of one Tx (green point) and 5 Rxs (red point), with LoS distances of 5~m, 10~m, 20~m, 40~m, 80~m, respectively. Tx is equipped at the top of a building of height 30~m. 
	The propagation paths with gain larger than -160~dB are plotted.  Particularly, in THz multi-path propagation, it is a common practice to omit the paths that are too weak, which contribute negligibly to the received signal power. 
		In our previous work [7], we omit the paths whose path gain is weaker than the strongest path by 27.8 dB. In this work, we set the path gain threshold at -160~dB to guarantee a 55~dB dynamic range, which guarantees to capture sufficient multi-path propagation.
	To illustrate the THz channel characteristic, we depict the DoA and path gain of the propagation paths in Fig~\ref{fig_WI_angles}. 
	The THz channel is extremely sparse, in which the number of available paths is 8. 
	This is caused by the high scattering
	and diffraction losses in the THz band. 
	Therefore, the number of channel parameters is much smaller than the dimension of the channel matrix in THz UM-MIMO systems, and performing channel parameter estimation is more appealing in the THz band than directly estimate the channel matrix. 
	
	To obtain the UM-MIMO channel matrix, we deploy different sizes of the UM-MIMO.
	Both Tx and Rx are equipped 4 subarrays, namely, $K_t = K_r= 4$, which are selected based on the sparse characteristic of the THz channel~\cite{ref_hybrid_magazine}. We deploy isotropic antennas with antenna gain equals to 0 dB. The number of antennas in each subarray is 256, 64 or 16, leading that the UM-MIMO contains 1024, 256 or 64 antennas at Tx and Rx, respectively. The antenna spacing within a subarray $d=\lambda/2$, while the distance between the subarrays $d_{tx}^{k_t},d_{tz}^{k_t},d_{rx}^{k_r},d_{rz}^{k_r}$, and the carrier frequency $f$ are changed as needed.
	Moreover, the spherical-wave channel is directly generated with MIMO antenna in Wireless InSite.
	To obtain the planar-wave channel, we deploy a single antenna at both Tx and Rx, and record the ray-tracing results in Wireless InSite, including the DoA and DoD, the delays, the phase, and the propagation distances of the arrival rays. The path gains are calculated using the THz path gain model in~\cite{Multi_ray_channel}, based on which we construct the planar-wave channel as~\eqref{equ_P_model}. 
	Moreover, the required channel parameters to construct the HSPM channel model in~\eqref{equ_HSPM_model} are generated by the ray-tracing results of Wireless Insite.
	All the numerical results are implemented on a PC with Intel(R) Xeon(R) CPU E5-2690 v4 @ 2.60 GHz and an Nvidia GeForce RTX 2080 Ti GPU.
	The DL-based methods including DCNN and existing solutions in the literature are carried out by using the PyCharm framework.
	
	\subsubsection{Training and testing of DCNN}
	The training data set is generated by simulation. Specifically, we randomly select 1000 Rx points in the Wireless InSite environment in~Fig.~\ref{fig_Simulation_environment}. The channel parameters of the planar array are recorded to compose the training labels. 
	We change the carrier frequencies as $f=0.2, 0.4$, and $0.8$~THz, respectively, to obtain 3000 HSPM matrices of different frequencies.
	For each channel matrix, the training process is conducted through the operations in~\eqref{Received_signal} and~\eqref{equ_Channel_observation}, to obtain the channel observation. Each codebook contains 4 codewords, in which the phase shift coefficient $\tilde{\omega}_{n_{k_r}^a,k_r}$ in~\eqref{equ_phase_shift_coefficient} is generated randomly, following uniform distribution in $[0,1]$. 
	In this way, the dimension of the input of the network equals to $16\times16\times3$.
	Moreover, AWGN noise is added for improved processing capability, which makes the received SNR equals to -20, -10, 0, and 10~dB. Therefore, the training data set is composed of 12000 samples in total.
	During our implementation, we consider the number of multi-path is known, which is a common practice in the literature~\cite{ref_EM_Bayesian}. 
		To form the testing data set, we select another 100 Rx points with SNR from ranging from -10 to 10~dB.
		In addition, since the characteristics of the THz channels are different at different frequencies, it requires enlarging the training data set to all frequencies to make the trained DCNN suitable for all carrier frequencies.

	\subsection{Accuracy of HSPM Channel model}
	\label{subsec_Accuracy of HSPM Channel model}
	\begin{figure}[t]
		\setlength{\belowcaptionskip}{0pt}
		\centering
		\subfigure[Approximation error versus communication distance.]{
			\includegraphics[width=0.31\textwidth]{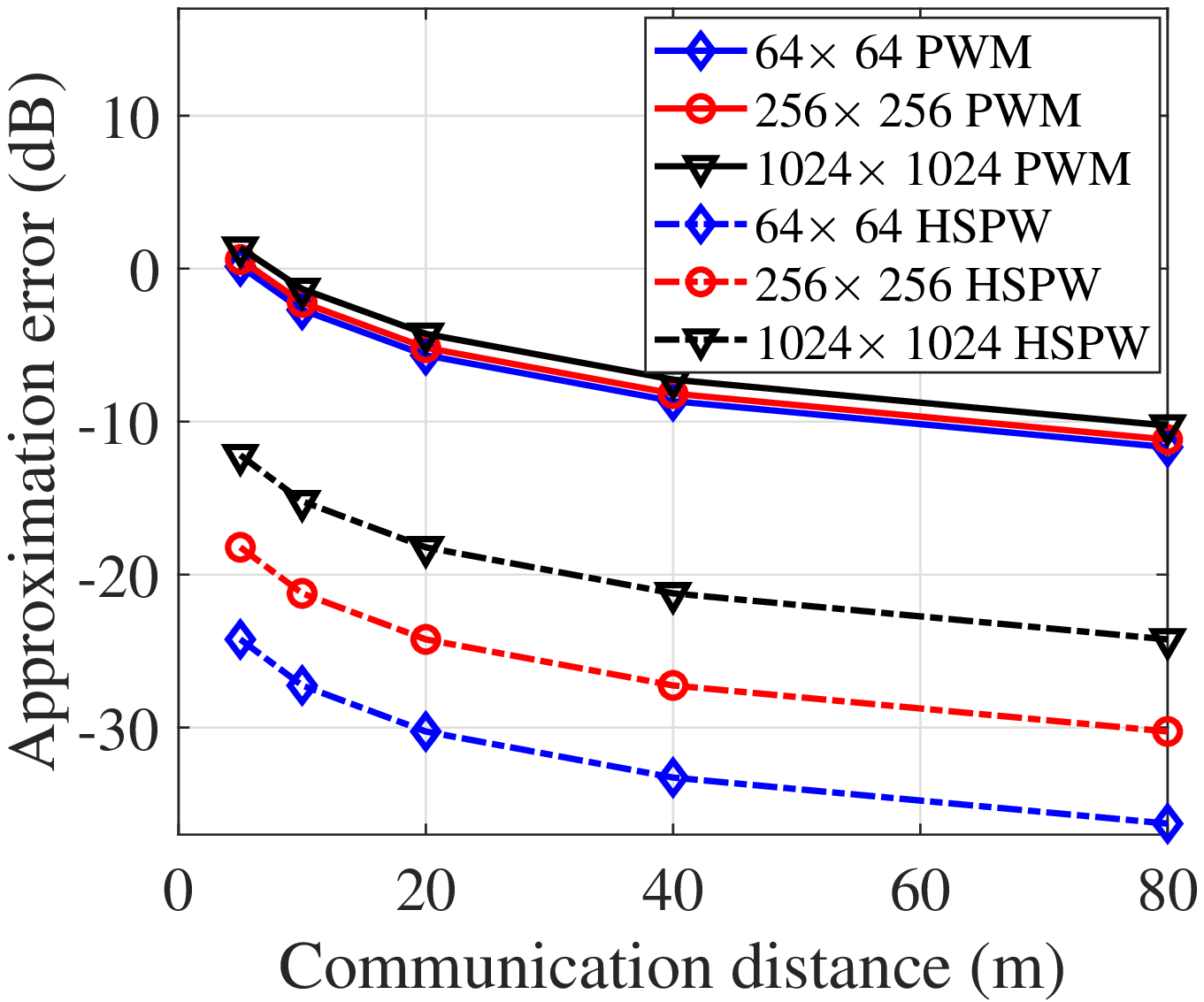} \label{fig_dist}}
		\subfigure[Approximation error versus subarray spacing.]{
			\includegraphics[width=0.31\textwidth]{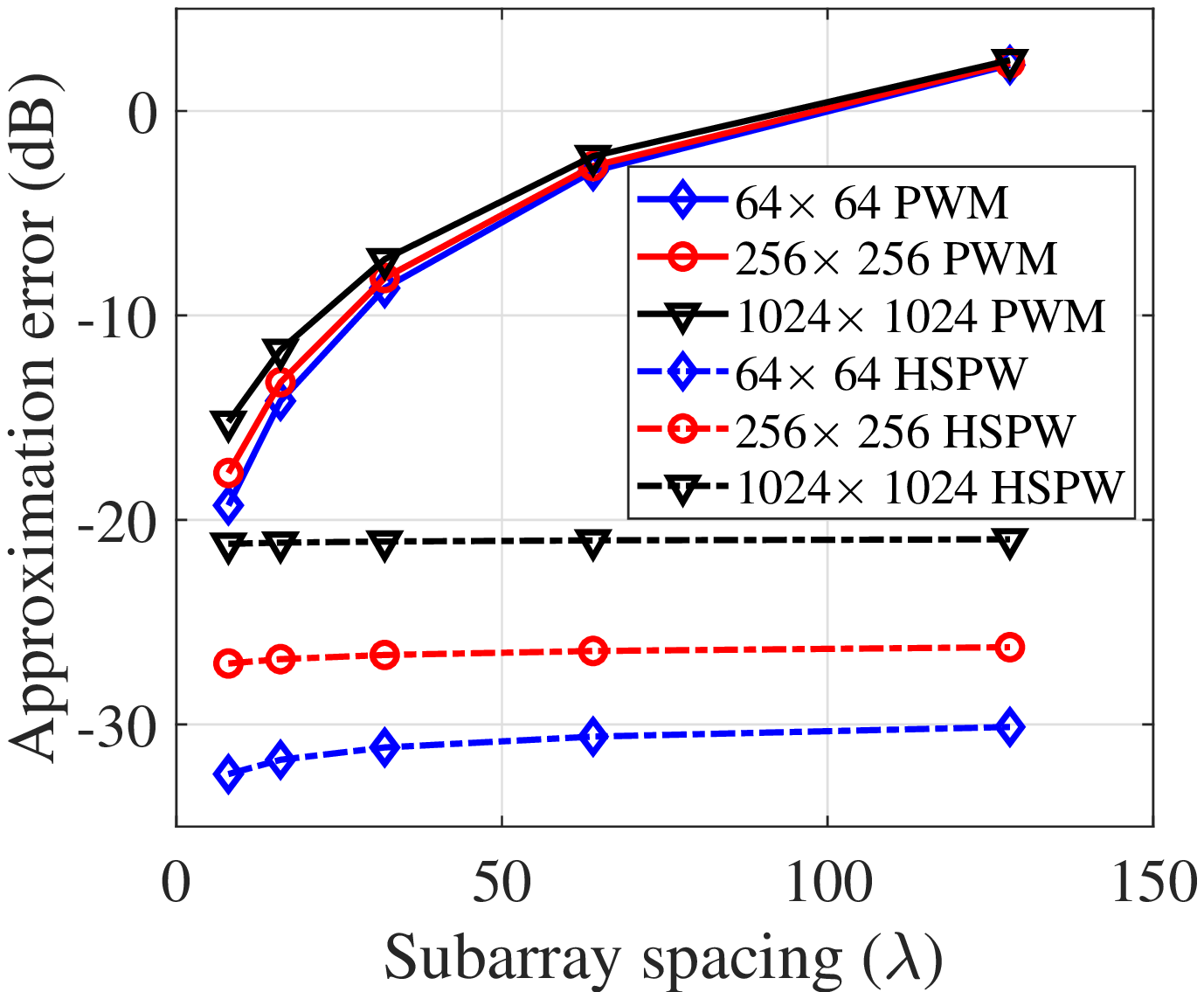} \label{fig_size}}
		\subfigure[Approximation error versus carrier frequancy.]{
			\includegraphics[width=0.31\textwidth]{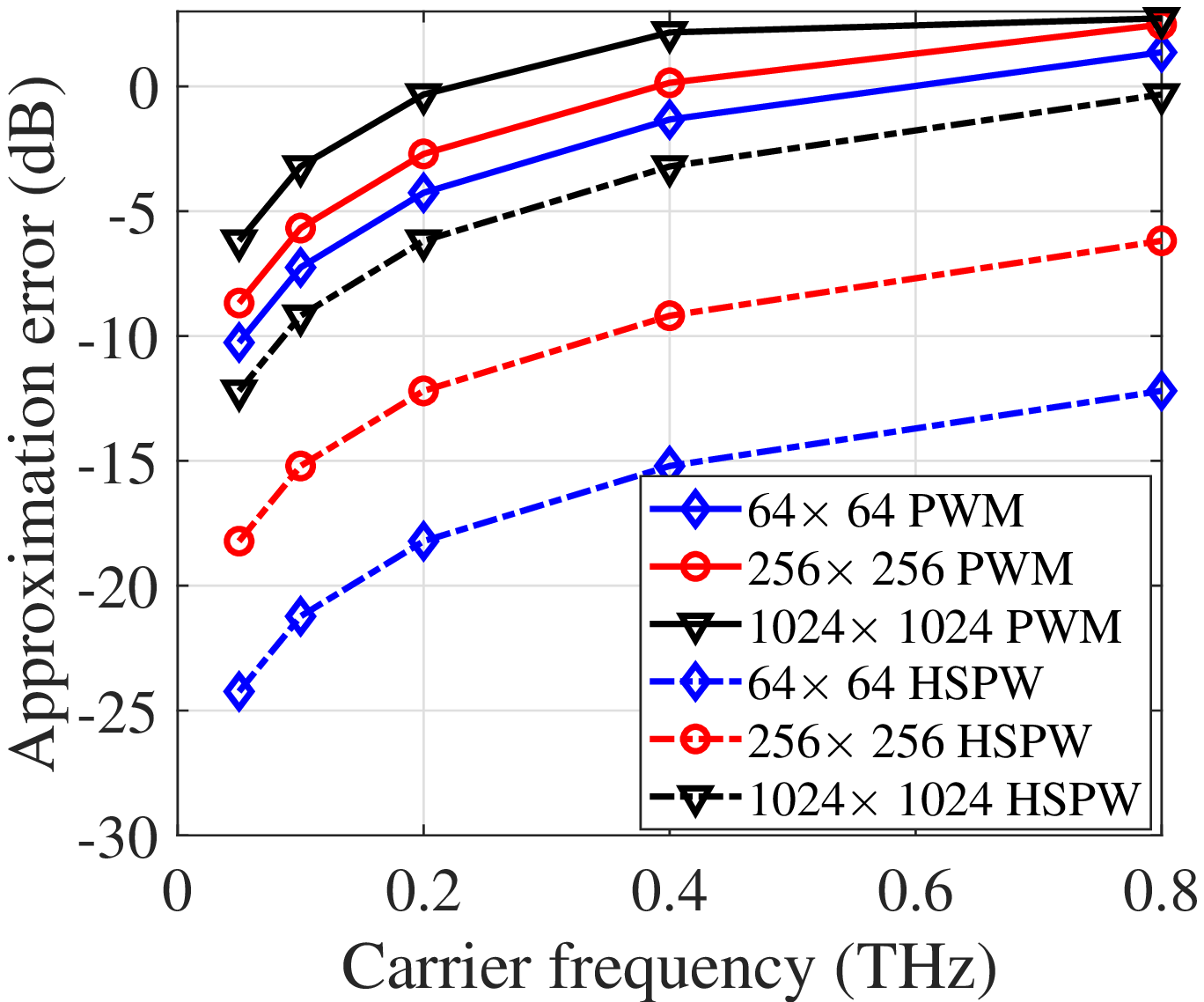} \label{fig_freq}}
		\caption{The accuracy of the PWM and HSPM.}
		\label{fig_HSPM_Accuracy}
		\vspace{-6mm}
	\end{figure}
	
	We begin by evaluating the accuracy of the proposed HSPM in~Fig.~\ref{fig_HSPM_Accuracy} by calculating the approximation error of the PWM and HSPM under different communication distances, subarray spacing, and carrier frequency, respectively.
	The approximation errors of the PWM and HSPM are obtained by calculating $\Vert\mathbf{H}_{\rm{P}}-\mathbf{H}_{\rm{S}}\Vert_{\rm{F}}/\Vert\mathbf{H}_{\rm{S}}\Vert_{\rm{F}}$ and $\Vert\mathbf{H}_{\rm{HSPM}}-\mathbf{H}_{\rm{S}}\Vert_{\rm{F}}/\Vert\mathbf{H}_{\rm{S}}\Vert_{\rm{F}}$, respectively. 
	First, the HSPM remains high accuracy in the THz band, in which the approximation error of the HSPM is much smaller than that of the PWM counterpart under different communication distances, subarray spacing, and carrier frequency. This is because the spherical-wave transmission is explored in the HSPM. 
	Specifically, as shown in Fig.~\ref{fig_dist}, with $f = 0.4$~THz, $d_{tx}^{k_t}=d_{tz}^{k_t}=d_{rx}^{k_r}=d_{rz}^{k_r} = 32\lambda$ and $N_t=N_r=1024$, the approximation error of the HSPM is 14~dB lower than the PWM channel at 20~m communication distance.	
	Moreover, the approximation errors of both PWM and HSPM decrease with the increment of communication distance, which confirms the result in Sec.~\ref{subsec_ccuracy Analysis for the Planar-wave Channel Model in THz UM-MIMO Systems.}. When $N_t=N_r=1024$, the approximation error of the PWM and HSPM decrease by 11.6~dB and 12~dB, respectively, as communication distance increases from 5~m to 80~m.

	As illustrated in Fig.~\ref{fig_size}, the effect of array size is explored with fixed transmission distance as 40~m. 
	The approximation error of the PWM increases with the increment of subarray spacing, which is consistent with the result in Sec.~\ref{subsec_ccuracy Analysis for the Planar-wave Channel Model in THz UM-MIMO Systems.}. Concretely, the approximation error of the PWM channel increases by 17.7~dB, when the subarray spacing increases from $8\lambda$ to $128\lambda$ and $N_t=N_r=1024$. 
	By contrast, the approximation error of HSPM remains almost unchanged under different subarray spacing. This is because the approximation error of the HSPM is mainly introduced by the planar-wave approximation in the subarray. By fixing the antenna spacing in the subarray as $\lambda/2$, the approximation error of the HSPM is stable. 
	Finally, with fixed array size and location of the antennas, we change the carrier frequency to obtain the result in Fig.~\ref{fig_freq}, where the communication distance equals to 40~m.
	The array size and the location of the antennas are obtained by setting $f=0.1$~THz, subarray spacing and antenna spacing in the subarray equal to 8$\lambda$ and $\lambda/2$, respectively. 
	The approximation errors of the HSPM and PWM increase with the carrier frequency. When $N_t=N_r=256$, the approximation error increases by 11.2~dB and 12.1~dB for the PWM and HSPM, respectively, when $f$ rises from 0.1~THz to 0.8~THz.
	
	\subsection{Performance of DCNN Channel Estimation}
	\label{subsec_Performance of DCNN Channel Estimation}

	\subsubsection{Convergence Evaluation}
	\begin{figure}[t]
		\setlength{\belowcaptionskip}{0pt}
		\centering
		\subfigure[Losses under different training epoch.]{
			\includegraphics[width=0.45\textwidth]{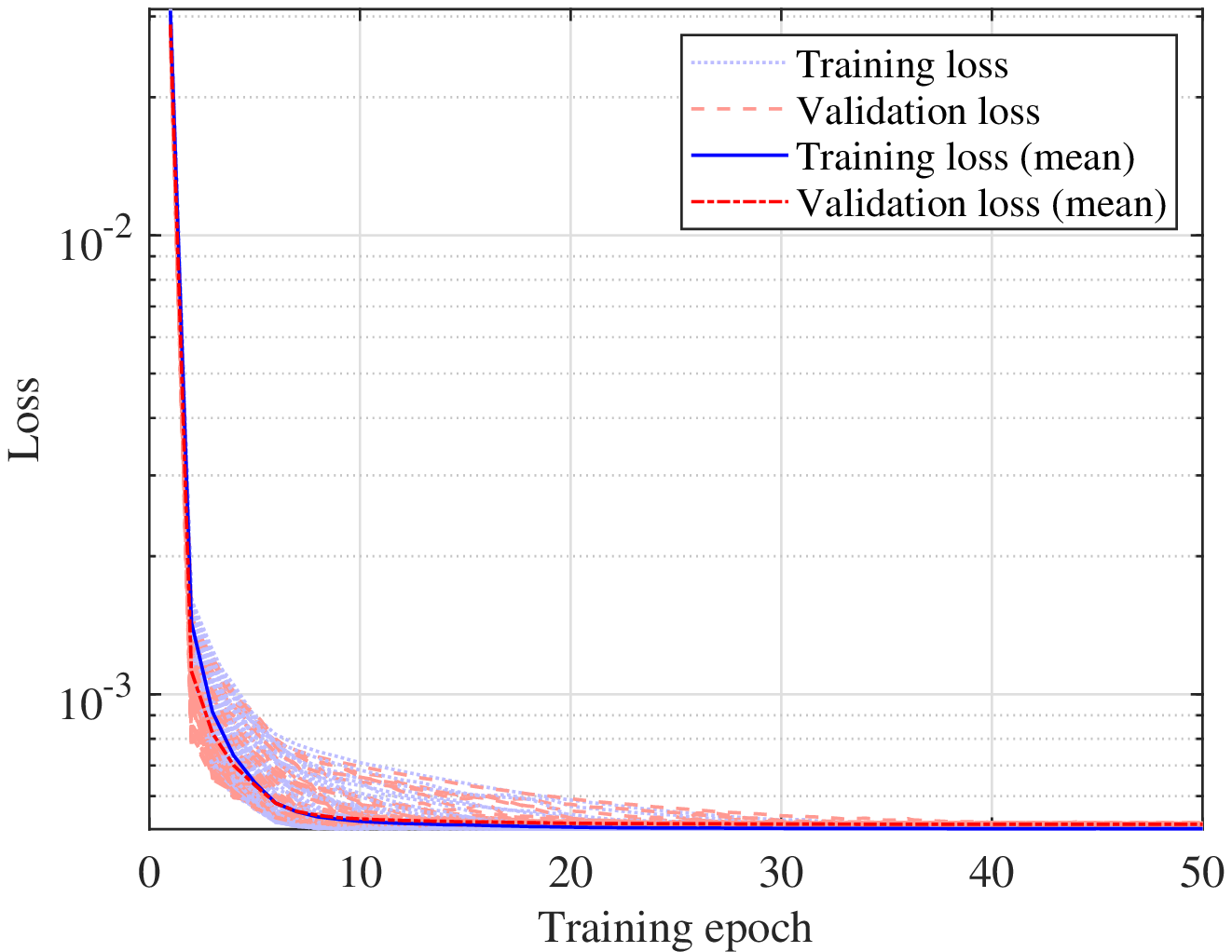} \label{fig_Loss_epoch}}
		\subfigure[Losses under different number of Rx points.]{
			\includegraphics[width=0.45\textwidth]{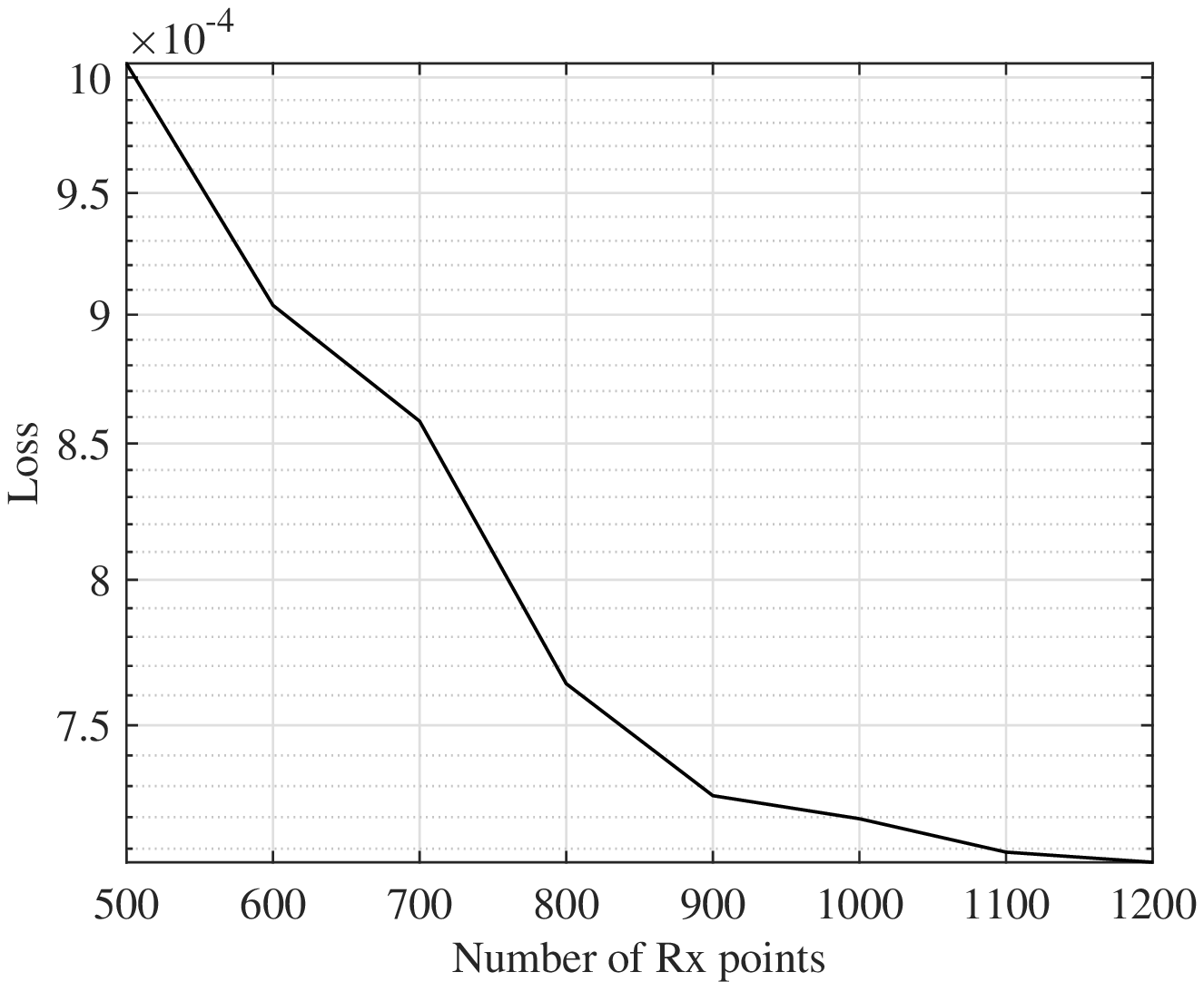} \label{fig_Loss_Rxpts}}
		\caption{The convergence performance of DCNN method.}
		\label{fig_Loss}
		\vspace{-6mm}
	\end{figure}
	
	The convergence performance of the proposed DCNN method is evaluated in Fig.~\ref{fig_Loss}, by analyzing the convergence speed and effect of the number of Rx points. 
	As shown in Fig.~\ref{fig_Loss_epoch}, we perform $100$ times network training and record the corresponding training and validation losses in the first 50 training epochs.
	The convergence performance is verified, as both training and validation losses tend to be stable after 40 epochs. 
	Furthermore, the mean values of training and validation losses for DCNN are close after network convergence, where the difference is on the order of $10^{-5}$ at the $40^{\rm{th}}$ epoch.

	Moreover, Fig.~\ref{fig_Loss_Rxpts} depicts the testing loss after 50 epochs under a various number of Rx points, which relates to different sizes of the training data set. The testing loss reduces as the number of Rx points grows, which suggests DCNN becomes more accurate with more training data. In addition, the value of testing loss tends to be saturated when the number of Rx points exceeds 900. 
	Compared to other DL-based CE methods~\cite{mmWave_CE_CNN} that require over 80000 samples to train the network, the proposed DCNN converges with only $15\%$ in the size of the training data set. This is owing to the fact that DCNN incorporates channel parameters as the training labels, leading that the output dimension is much smaller than that by choosing the channel matrix as the training label. Therefore, DCNN has a reduced network complexity and requires a significantly smaller size of the training data set.
	
	\subsubsection{Estimation Accuracy}\label{subsubsec_Estimation Accuracy}

\begin{figure}[t]
		\setlength{\belowcaptionskip}{0pt}
		\centering
		\subfigure[Estimation error of angle.]{
			\includegraphics[width=0.31\textwidth]{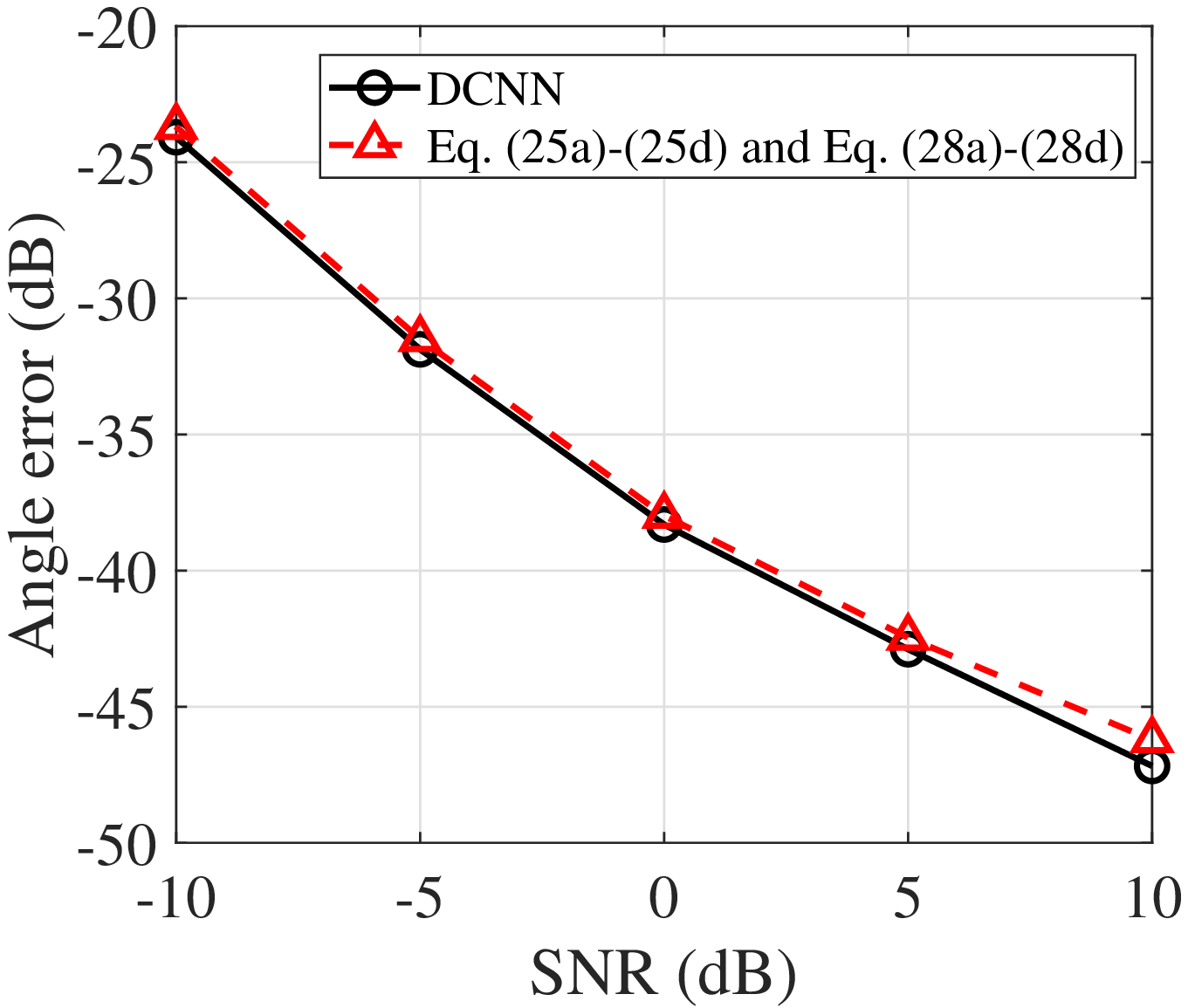} \label{fig_angle_err}}
		\subfigure[Estimation error of distance.]{
			\includegraphics[width=0.31\textwidth]{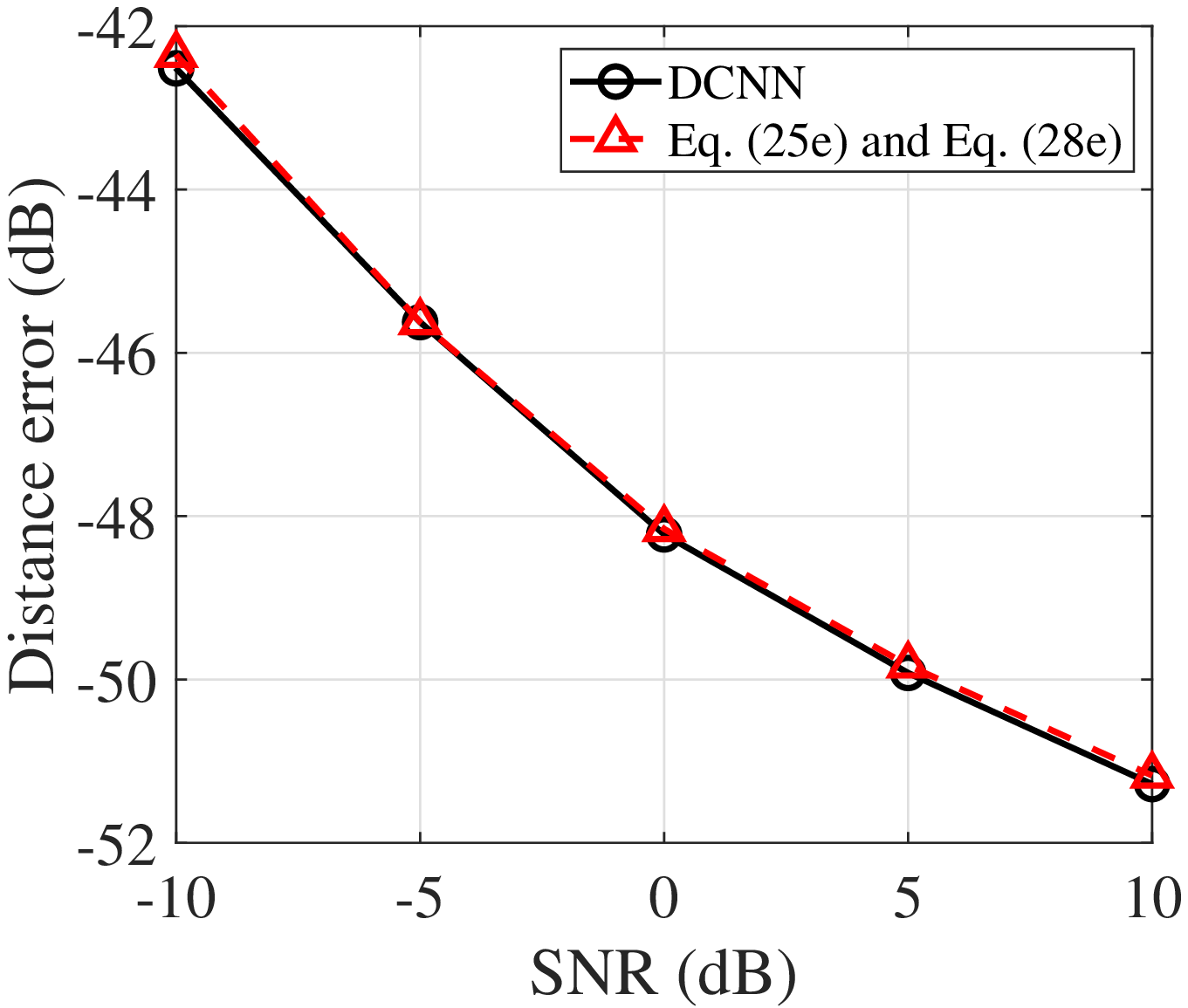} \label{fig_dist_err}}
		\subfigure[Estimation error of path gain.]{
			\includegraphics[width=0.31\textwidth]{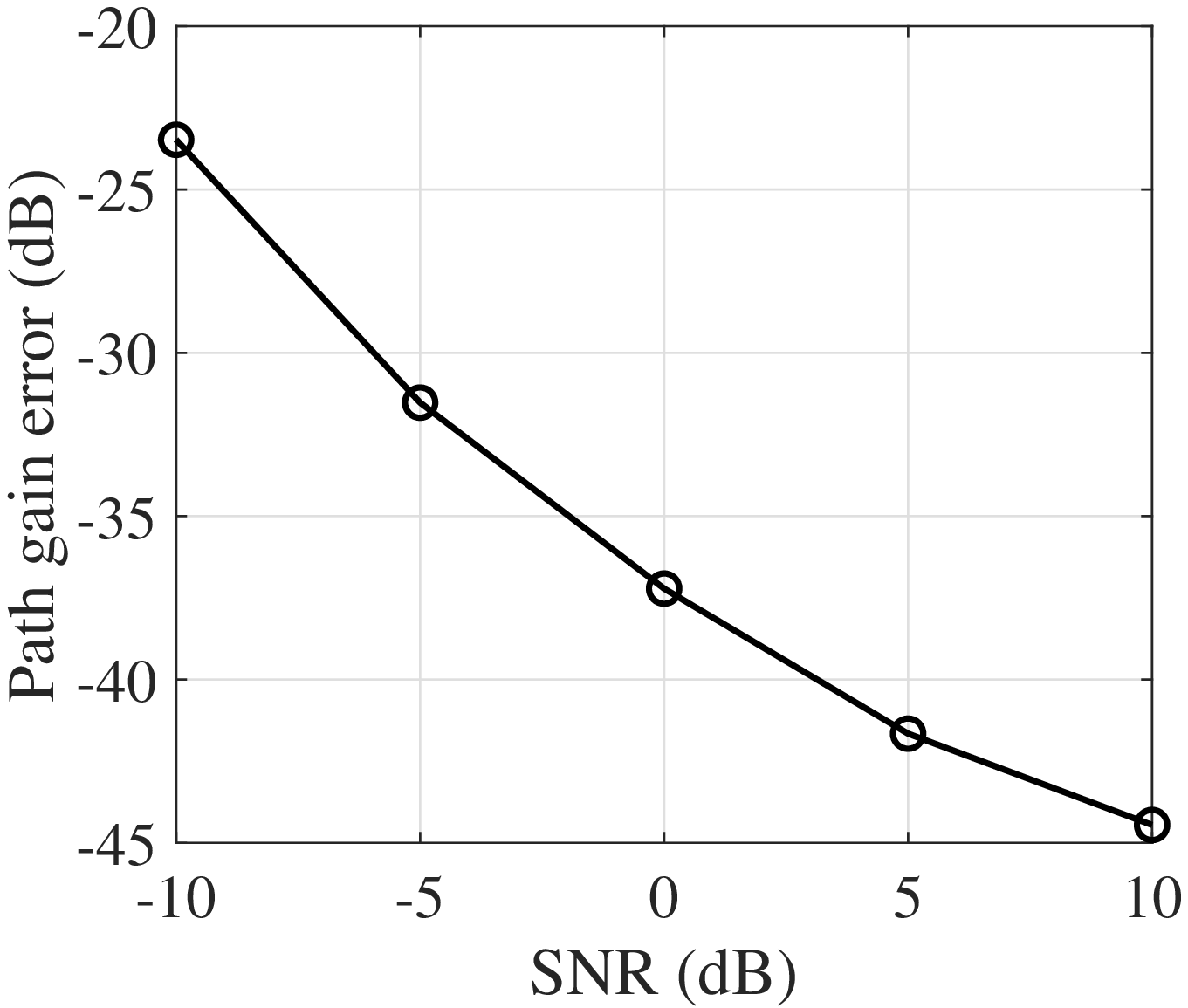} \label{fig_gain_err}}
		\caption{Channel parameter estimation accuracy of the DCNN method.}
		\label{fig_parameter_err}
		\vspace{-6mm}
	\end{figure}

	The estimation accuracy of the proposed DCNN method is evaluated in terms of the parameter estimation accuracy and the CE normalized-mean-square-error (NMSE). 
	The estimation errors of the angles, distance, and channel gain are calculated as $\mathbb{E} 
	\{\Arrowvert{\boldsymbol{\varsigma}}-\hat{\boldsymbol{\varsigma}}\Arrowvert/ \Arrowvert{\boldsymbol{\varsigma}}\Arrowvert \}$, $\mathbb{E}\{\Arrowvert\hat{D_{p}^{11}}-D_p^{11}\Arrowvert/\Arrowvert D_p^{11}\Arrowvert\}$ and $\mathbb{E}\{\Arrowvert\hat{|\alpha_{p}^{11}|}-|\alpha_{p}^{11}|\Arrowvert/\Arrowvert|\alpha_{p}^{11}|\Arrowvert\}$, respectively. Moreover, the NMSE of the CE result is defined as $\Arrowvert\hat{\mathbf{H}} - \mathbf{H}_{\rm HSPM}\Arrowvert_{\rm F}/ \Arrowvert\mathbf{H}_{\rm HSPM} \Arrowvert_{\rm F}$, where $\hat{\mathbf{H}}$ represents the estimated channel matrix. All the results are obtained by averaging for 5000 trials of Monte Carlo simulations.
	As depicted in Fig.~\ref{fig_parameter_err}, the channel parameter estimation error under different received SNR is investigated. Both DCNN and HSPM CE achieve substantial accuracy in estimating the channel parameters, where the estimation errors of the angles, distance, and path gain reach -38.2~dB, -48.8~dB, and -37.2~dB, respectively, at SNR$=0$~dB. 
	Moreover, due to cumulative error, the estimation error of the derived results in equations~\eqref{equ_los_para} and~\eqref{equ_nlos_para} in the second estimation phase is slightly higher than the DCNN method. As shown in Fig.~\ref{fig_angle_err}, when SNR$=$5dB, the estimation error of the derived result is 0.37~dB higher than the DCNN method.

\begin{figure}[t]
		\centering
		{\includegraphics[width=3.3in]{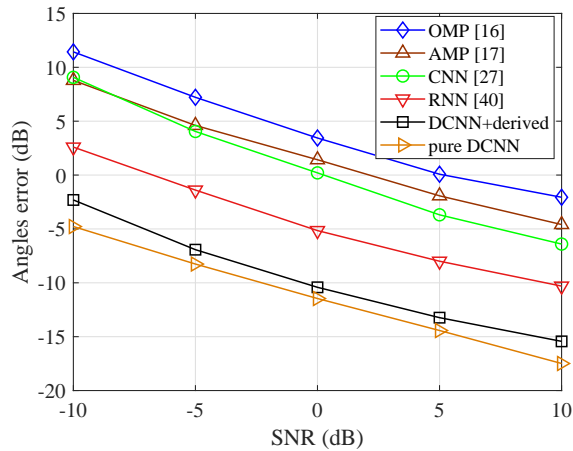}}
		\caption{NMSE performance comparison of different CE methods.} 
		\label{NMSE_Comp}
		\vspace{-5mm}
	\end{figure}

	In Fig.~\ref{NMSE_Comp}, we compare the NMSE performance of the DCNN method with two classical on-grid CS-based algorithms, including OMP~\cite{OMP} and AMP~\cite{AMP}, as well two off-grid DL-based schemes, namely, CNN~\cite{mmWave_CE_CNN}, recurrent neural network (RNN)~\cite{RNN}. The legend ``DCNN+derived'' denotes the proposed two-phase channel estimation, i.e., DCNN for the reference subarray and derivations for other subarrays. The legend ``pure DCNN'' represents the scheme which uses the DCNN for all subarrays.
	For on-grid solutions, the number of spatial grids equals to the number of antennas as $1024\times1024$.
	Since RNN requires data sequences as the input, the shape of the $16\times16\times3$ dimensional sample data is transformed into a matrix of $16\times48$, of which each row is the data sequence. 
	Moreover, the same data set is used among these algorithms for a fair comparison.
	The proposed DCNN method performs the best and achieves the lowest NMSE among the listed algorithms. Specifically, when SNR=0 dB, the NMSE of the DCNN is 6 dB lower than the RNN method. Although the two-phase DCNN estimation method experiences performance degradation due to the error propagation, in which the estimation NMSE is 1~dB lower than the DCNN method when SNR=0 dB, it still outperforms the listed algorithms. Specifically, when SNR=10 dB, the NMSE of the DCNN is 5.2~dB lower than the RNN method.
	Moreover, due to the consideration of spherical-wave propagation in the HSPM, traditional CS-based OMP and AMP suffer from the highest NMSE induced by the grid mismatch. By contrast, the estimation accuracy of all the DL-based solutions outperforms the traditional methods when the received SNR exceeds -5~dB. This suggests an improved performance by DL method.
	
	\begin{table}[t]
		\centering
		\caption{Comparison on computational complexity and running time for different methods.}
		\begin{tabular}{ccc}
			\toprule
			\textbf{Method}&\textbf{Computational Complexity}& \textbf{Running Time (ms)}\\
			\midrule
			OMP~\cite{OMP} &$\mathcal{O}\left( (N_pN_t)^2\right)$ &$221$\\
			AMP~\cite{AMP}&$\mathcal{O}\left( (N_pN_t)^2\right)$&$372$\\
			CNN~\cite{mmWave_CE_CNN} & $\mathcal{O}\left( b(N_tN_r)^2\right)$ &$3.64$\\
			RNN~\cite{RNN}&$\mathcal{O}\left(c C^3K_tK_r\right)$&$0.085$\\
			DCNN &$\mathcal{O}\left( b(C^2K_tK_r)^2\right)$ &$0.172$\\
			\bottomrule
		\end{tabular}
		\label{Table_complexity}
		\vspace{-6mm}
	\end{table}
	
	\subsubsection{Computational Complexity}\label{subsubsec_Computational Complexity}

	Finally, the computational complexity and running time of the proposed DCNN method and literature solutions are compared in TABLE~\ref{Table_complexity}, where $b=3$ denotes the dimension of the CV layer in CNN~\cite{mmWave_CE_CNN} and DCNN, while in RNN~\cite{RNN}, $c=3$ represents the number of input channels.
	We observe that RNN processes the lowest computational complexity and running time, which comes at the cost of high estimation error. 
	Moreover, a low computational complexity is achieved by the proposed DCNN method, by which the running time reach 0.172~ms.
	This is attributed to the superiority of the designed DCNN architecture. 
	In particular, the inserted MP layers are helpful to reduce the required number of network parameters, which yet remain the efficiency for extracting the features of the channel. 
	Apart from that, it is noteworthy that the dimension of the channel matrix in THz UM-MIMO systems is large, which induces significantly high complexity of the existing OMP, AMP, and CNN methods, whose running times are 221~ms, 372~ms, and 3.64~ms, respectively.

	\section{Conclusion}\label{sec_Conclusion} 
	In this paper, we have evaluated the HSPM channel model and proposed a two-phase CE mechanism for THz UM-MIMO systems.
	First, we analytically derive the closed-form expression of the approximation error between the PWM and SWM for the 2D planar array in THz UM-MIMO systems. 
	By exploiting the spherical-wave propagation and parameter shift among subarrays, the HSPM is investigated.
		Extensive comparisons confirm that the HSPM uses a small number of channel parameters including the azimuth and elevation angles of departure and arrival, the amplitude of the path gain, and the communication distances to achieve high accuracy.
	The proposed two-phase CE first trains a DCNN network to learn the estimation of the channel parameters at the reference subarrays. 
	Then, based on the estimated channel parameters of the reference subarray, we derive the expression of the channel parameters using the geometric relationships and reconstruct the channel matrix to complete the CE process. 
	
	To evaluate the performance of the proposed HSPM and HSPM CE mechanism, respectively, we deployed a ray-tracing tool to provide extensive numerical results. 
	The accuracy of the HSPM is confirmed by comparing the modeling error with the PWM under various communication distances, subarray spacings, and carrier frequencies. 
	While the performance of the HSPM CE mechanism is revealed in terms of convergence performance, estimation accuracy, and computational complexity.
	The HSPM achieves similar accuracy with the SWM and is more accurate than the PWM channel with different communication distances, array sizes as well as carrier frequencies.
	The HSPM achieves 14~dB higher accuracy than PWM at 20m communication distance when $N_t=N_r=1024$, and subarray spacing equals $32\lambda$.
	Compared to the existing CE algorithms, the designed DCNN network convergence fast and achieves high-resolution CE with substantially reduced complexity, whose estimation accuracy is improved by 5.2~dB in 0.172~ms.
	\begin{appendices}
		\section{Derivation of the approximation error}\label{Sec_Appendix}

	The approximation error $\epsilon_{il}$ in \eqref{equ_error_planar_2} can be rewritten as
	\begin{equation}\label{equ_appendix_1}
	\begin{split}
	\epsilon_{il}&\approx \left\lvert2{\rm sin}\left(\frac{\pi}{\lambda}(D^{il}-D^{11}-\Delta D^{il}\right)\right\rvert.
	\end{split}
	\end{equation}
	To calculate~\eqref{equ_appendix_1}, we consider the coordinate system at Tx of the reference antenna. As shown in Fig.~\ref{fig_system_model}, the vectors $\overrightarrow{v_{t,l}} $ direct from the reference antenna to the $ l^{\rm th}$ antenna at Tx, $\overrightarrow{u_{t}}$ direct from the reference antenna at Tx to the reference antenna at Rx, $\overrightarrow{v_{r,i}}$ direct from the reference antenna to the $ i^{\rm th}$ antenna at Rx, respectively. The coordinates are represented as 
	$
	\overrightarrow{v_{t,l}} = ((m_{k_tx} + n_{k_tx})d, 0, -(m_{k_tz} + n_{k_tz})d)$, 	$
	\overrightarrow{u_{t}}= ({\rm sin}\theta^{11}_{t1}{\rm cos}\phi^{11}_{t1},$ $ {\rm cos}\theta^{11}_{t1}{\rm cos}\phi^{11}_{t1}$ and ${\rm sin}\phi^{11}_{t1}),
	\overrightarrow{v_{r,i}} = ((m_{k_rx}\! + n_{k_rx})d, 0, -(m_{k_rz} \!+ n_{k_rz})d)$, respectively. 	
	Then, $D^{il}-D^{11}-\Delta D^{il}$ in~(13) is derived as
	\begin{subequations}
		\begin{align}
		D^{il}-D^{11}-\Delta D^{il} &= \Vert -\overrightarrow{v_{t,l}} + D^{11} \overrightarrow{u_{t}} + \overrightarrow{v_{r,i}} \Vert_2 - D^{11}-\Delta D^{il},\\
		&= D^{11}\left( \underbrace{\sqrt{1+\frac{2(\overrightarrow{v_{r,l}} \overrightarrow{u_{t}} - \overrightarrow{v_{t,l}} \overrightarrow{u_{t}} )}{D^{11}} + \frac{\Vert\overrightarrow{v_{r,i}} - \overrightarrow{v_{t,l}}\Vert_2^2 }{(D^{11})^2}}}_{(**)} - 1\right)-\Delta D^{il}.
		\end{align}
	\end{subequations}
	By applying the Taylor expansion to $(**)$, we obtain
	\begin{subequations}\label{equ_appendix_2}
		\begin{align}
		&D^{il}-D^{11}-\Delta D^{il}\notag
		\\=&~\overrightarrow{v_{r,i}} \overrightarrow{u_{t}} - \overrightarrow{v_{t,l}} \overrightarrow{u_{t}} -\Delta D^{il}+
		\frac{1}{2D^{11}} 
		\left\{
		\Vert\overrightarrow{v_{r,i}} -
		\overrightarrow{v_{t,l}}\Vert_2^2- \left(\overrightarrow{v_{r,i}} \overrightarrow{u_{t}} - \overrightarrow{v_{t,l}} \overrightarrow{u_{t}}\right)^2\right\}
		+\mathcal{P},\\
		=&~
		\frac{1}{2D^{11}} 
		\left[ (\overrightarrow{v_{r,i}})^2 -2\overrightarrow{v_{r,i}}\overrightarrow{v_{t,l}} +(\overrightarrow{v_{t,l}})^2 -\left( \overrightarrow{v_{r,i}} \overrightarrow{u_{t}}\right)^2
		+ 2\overrightarrow{v_{r,i}} \overrightarrow{u_{t}} \overrightarrow{v_{t,l}} \overrightarrow{u_{t}}- \left( \overrightarrow{v_{t,l}} \overrightarrow{u_{t}}\right)^2
		\right]
		+\mathcal{P},
		\\
		=&~\notag
		\frac{d^2}{2D^{11}} \Big\{(m_{k_rx} + n_{k_rx})^2 + (m_{k_rz} + n_{k_rz})^2 + ((m_{k_tx} + n_{k_tx}))^2 + (m_{k_tz} + n_{k_tz})^2 \notag
		\\&\notag-2(m_{k_tx} + n_{k_tx})(m_{k_rx} + n_{k_rx}) \notag-2
		(m_{k_tz} + n_{k_tz})(m_{k_rz} + n_{k_rz}) \\&\notag- \left[(m_{k_rz} + n_{k_rz}){\rm sin}\theta^{11}_{t1}{\rm cos}\phi^{11}_{t1} - (m_{k_rz} + n_{k_rz}){\rm sin}\phi^{11}_{t1} \right]^2\\\notag
		& +2 \left[\!(m_{k_rz} + n_{k_rz}){\rm sin}\theta^{11}_{t1}{\rm cos}\phi^{11}_{t1} \!-\! (m_{k_rz} + n_{k_rz}){\rm sin}\phi^{11}_{t1} \!\right]
		\\& \notag \times
		\left[(m_{k_tx} + n_{k_tx}){\rm sin}\theta^{11}_{t1}{\rm cos}\phi^{11}_{t1} - (m_{k_tz} + n_{k_tz}){\rm sin}\phi^{11}_{t1} \right]\notag
		\\&
		-\left[\!((m_{k_tx} + n_{k_tx})){\rm sin}\theta^{11}_{t1}{\rm cos}\phi^{11}_{t1} - (m_{k_tz} + n_{k_tz}){\rm sin}\phi^{11}_{t1} \right]^2\Big\} + \mathcal{P},
		\\ =& ~ \frac{d^2 }{2D^{11}} \left[(m_{k_rx}-m_{k_tx} + n_{k_rx} - n_{k_tx})^2
		\left({\rm sin}^2\theta^{11}_{t1} {\rm cos}^2\phi^{11}_{t1} + {\rm cos}^2\theta^{11}_{t1}\right) \right.
		\notag
		\\&\left.+ (m_{k_rz}-m_{k_tz} + n_{k_rz} - n_{k_tz})^2{\rm cos}^2\phi^{11}_{t1}
		\right] + \mathcal{P},
		\end{align}
	\end{subequations}
	where $\overrightarrow{v_{r,i}} \overrightarrow{u_{t}} - \overrightarrow{v_{t,l}} \overrightarrow{u_{t}} =\Delta D^{il}$, and $\mathcal{P}$ represents the remainder of Taylor expansion of more than three orders, whose impact is far less than the first and second order terms and could be omitted. 
	Based on~\eqref{equ_appendix_2}, we can obtain the result in~\eqref{equ_approximation_error}.

	\end{appendices}
	
	\bibliographystyle{IEEEtran}
	\bibliography{reference} 
\end{document}